\pdfoutput=1
\documentclass[cernpreprint,texlive=2011,txfonts,UKenglish,texmf]{atlasdoc}

\usepackage{atlaspackage}
\usepackage{atlasbiblatex}

\usepackage{atlasphysics}

\graphicspath{{logos/}{figures/}}

\usepackage{hhpaper-defs}
\usepackage{multirow}

\hypersetup{pdftitle={ATLAS draft},pdfauthor={The ATLAS Collaboration}}

\AtlasTitle{Searches for Higgs boson pair production in the $hh\!\to\! \bb\tau\tau,\, \gamma\gamma WW^*,\, \gamma\gamma \bb,\, \bb\bb$ channels with the ATLAS detector}

\author{The ATLAS Collaboration}

\AtlasRefCode{Higgs-2013-33}

\PreprintIdNumber{CERN-PH-EP-2015-225}

\AtlasJournalRef{Phys. Rev. D92, 092004 (2015)}
\AtlasDOI{10.1103/PhysRevD.92.092004}

\AtlasAbstract{%
Searches for both resonant and nonresonant Higgs boson pair production are performed in the $hh\!\to\! \bb\tau\tau,\, \gamma\gamma WW^*$ final states using 20.3~\ifb\ of $pp$ collision data at a center-of-mass energy of 8 TeV recorded with the ATLAS detector at the Large Hadron Collider. No evidence of their production is observed and 95\% confidence-level upper limits on the production cross sections are set. These results are then combined with the published results of the $hh\!\to\! \gamma\gamma \bb,\, \bb\bb$ analyses. An upper limit of 0.69~(0.47)~pb on the nonresonant $hh$ production is observed (expected), corresponding to 70 (48) times the SM $gg\!\to\! hh$ cross section. For production via narrow resonances, cross-section limits of $hh$ production from a heavy Higgs boson decay are set as a function of the heavy Higgs boson mass. The observed (expected) limits range from 2.1 (1.1)~pb at 260~GeV to 0.011~(0.018)~pb at 1000~GeV. These results are interpreted in the context of two simplified scenarios of the Minimal Supersymmetric Standard Model. 
}

\AtlasCoverSupportingNote{$\bbtt$}{https://cds.cern.ch/record/1967500}
\AtlasCoverSupportingNote{$\yyWW$}{https://cds.cern.ch/record/1967498}
\AtlasCoverSupportingNote{Combination}{https://cds.cern.ch/record/1984111}

\AtlasCoverCommentsDeadline{25 August 2015}

\AtlasCoverAnalysisTeam{Jahred~Adelman, Andrew~Daniells, Yaquan~Fang~(*), Ryosuke~Fuchi, Keita~Hanawa~(*), Shan~Jin, Nikos~Konstantinidis, Qi~Li, Xinchou~Lou, Allison~McCarn, Konstantinos~Nikolopoulos, Jianming~Qian~(*), Nikolaos~Rompotis, James~Saxon, Xiaohu~Sun, David~Wardrope, Huijun~Zhang}

 \AtlasCoverEdBoardMember{Bill Murray~(*), Aaron Armbruster, Stanley Lai, Magdalena Slawinska, Wei-ming Yao}

\AtlasCoverEgroupEditors{atlas-higg-2013-33-editors@cern.ch}

\AtlasCoverEgroupEdBoard{atlas-higg-2013-33-editorial-board@cern.ch}

\begin{document}

\newpage

\section{Introduction}
\label{sec:introduction}

\begin{figure}[htb]
\begin{center}
\subfloat[]{\includegraphics[width=0.45\textwidth]{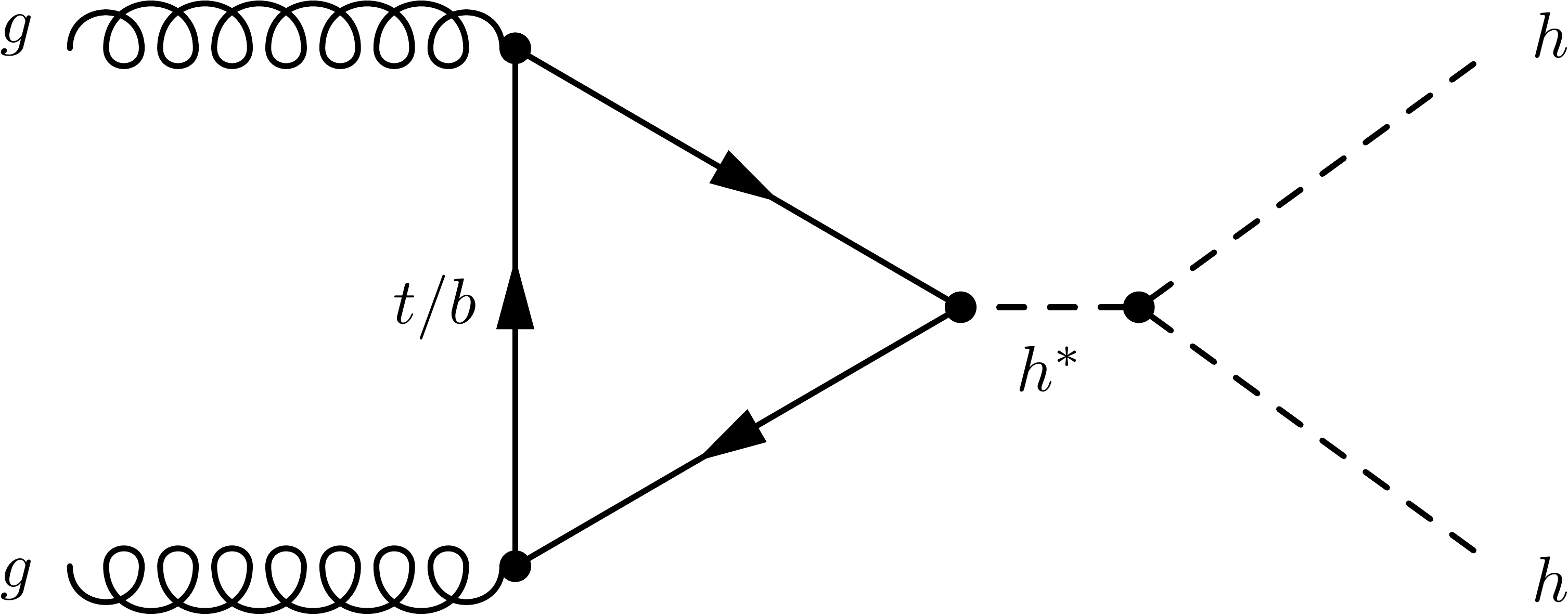}}\hspace*{1.0cm}
\subfloat[]{\includegraphics[width=0.45\textwidth]{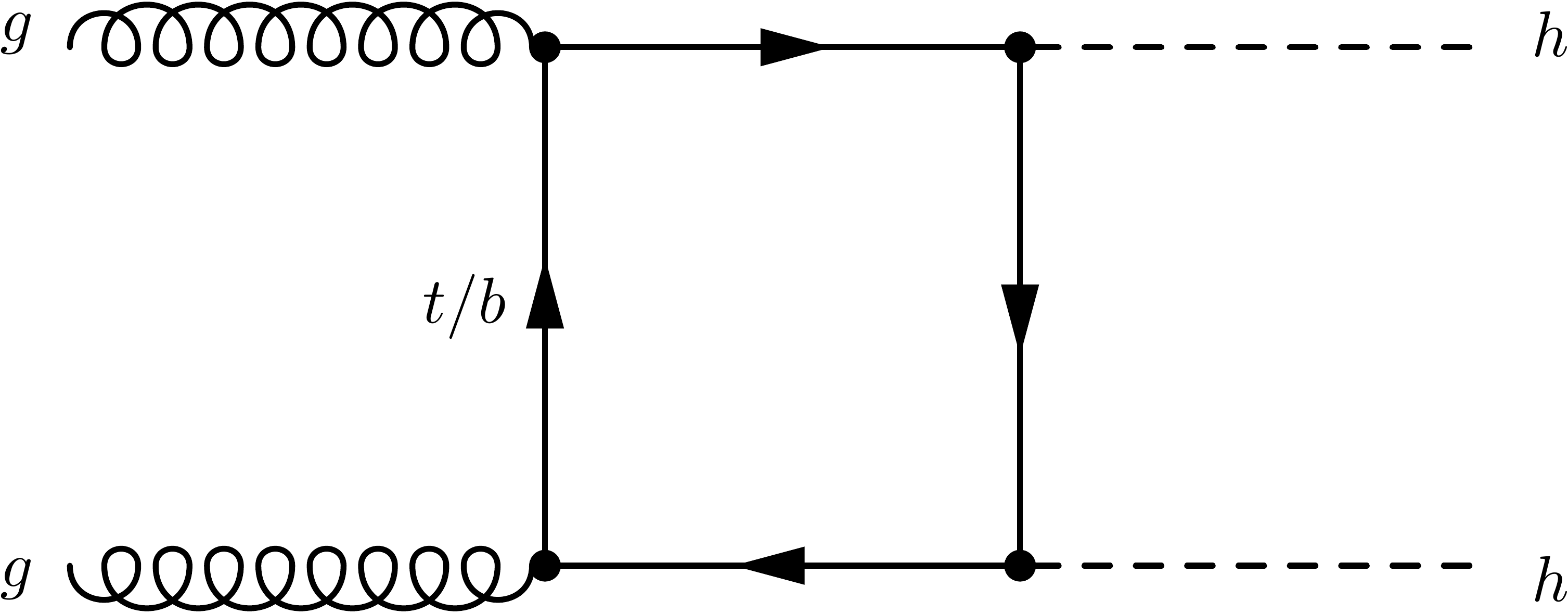}}
\end{center}
\caption{Leading-order Feynman diagrams of the nonresonant production of Higgs boson pairs in the Standard Model through (a) the Higgs boson self-coupling and (b) the Higgs--fermion Yukawa interactions.}
\label{fig:gghh}
\end{figure}

The Higgs boson discovered at the LHC in 2012~\cite{atlas:2012obs,cms:2012obs} opens a window for testing the scalar sector of the Standard Model~(SM) and its possible extensions. Since the discovery, significant progress has been made in measuring its coupling strengths to fermions and vector bosons~\cite{Aad:2013wqa,Khachatryan:2014jba,Aad:2014xzb,Aad:2015vsa} as well as in studying its spin and its charge-conjugate and parity~(CP) properties~\cite{Aad:2013xqa,Khachatryan:2014kca}. All results are consistent with those expected for the SM Higgs boson (here denoted by $h$). Within the SM, the existence of the Higgs boson is a consequence of the electroweak symmetry breaking~(EWSB). This also predicts self-coupling between Higgs bosons, the measurement of which is crucial in testing the mechanism of EWSB. The self-coupling is one mechanism for Higgs boson pair production as shown in Fig.~\ref{fig:gghh}(a). Higgs boson pairs can also be produced through other interactions such as the Higgs--fermion Yukawa interactions (Fig.~\ref{fig:gghh}(b)) in the Standard Model. These processes are collectively referred to as nonresonant production in this paper.  

Higgs boson pair production at the LHC as a probe of the self-coupling has been extensively studied in the literature~\cite{Baur:2002qd,Baur:2003gp,Dolan:2012rv,Baglio:2012np,Lu:2015qqa}. One conclusion~\cite{Dawson:2013bba} is that the data collected so far (approximately 25~\ifb\ in total) are insensitive to the self-coupling in the SM,  because of the expected small signal rates~\cite{Dawson:1998py,Grigo:2013rya,deFlorian:2013jea} and large backgrounds. However, it is essential to quantify the sensitivity of the current dataset and to develop tools for future measurements. Moreover, physics beyond the Standard Model~(BSM) can potentially enhance the production rate and alter the event kinematics. 
For example, in the Minimal Supersymmetric Standard Model~(MSSM)~\cite{Haber:1984rc}, a heavy CP-even neutral Higgs boson $H$ can decay to a pair of lighter Higgs bosons. Production of $H$ followed by its decay $H\!\to\! hh$ would lead to a new resonant process of Higgs boson pair production, in contrast to the nonresonant $hh$ production predicted by the SM~(Fig.~\ref{fig:gghh}). In composite Higgs models such as those discussed in Refs.~\cite{Grober:2010yv,Contino:2012xk}, increased production of nonresonant Higgs boson pairs is also expected.  

Both the ATLAS and CMS collaborations have searched for nonresonant
and/or resonant Higgs boson pair
production~\cite{Aad:2014yja,Aad:2015uka,Khachatryan:2015yea}. In
particular, ATLAS has published the results of searches in the
$\bbyy$~\cite{Aad:2014yja} and $\bbbb$~\cite{Aad:2015uka} decay channels.\footnote{Notations indicating particle charges or antiparticles are generally omitted throughout this paper.}  
In this paper, searches in two additional $hh$ decay final
states, $\bb \tau\tau$ and $\gamma\gamma WW^*$, are reported.  For the
$\bbtt$ analysis, one tau lepton is required to decay to an electron
 or a muon, collectively referred to as $\ell$, and the
other tau lepton decays to hadrons ($\tauhad$).  For $\yyWW$, the $h\!\to\! WW^*\!\to\! \ell\nu qq'$
decay signature is considered in this study. The results of these new analyses are combined with the published results of $\bbyy$ and $\bbbb$ for both nonresonant and resonant production. The resonance mass $m_H$ considered in this paper ranges from 260~GeV to 1000~GeV. The lower bound is dictated by the $2m_h$ threshold while the upper bound is set by the search range of the $\bbtt$ analysis. 
The light Higgs boson mass $m_h$ is assumed to be $125.4$~GeV, the central value of the ATLAS measurement~\cite{Aad:2014aba}. At this mass value, the SM predictions~\cite{Dittmaier:2011ti,Dittmaier:2012vm,Heinemeyer:2013tqa} for the decay fractions of $hh\!\to\!\bb\bb,\,\bb\tau\tau,\,\bb\gamma\gamma$ and $\gamma\gamma WW^*$ are, respectively, 32.6\%, 7.1\%, 0.26\% and 0.10\%. 
The resonant search assumes that gluon fusion is the production mechanism for a heavy Higgs boson that can subsequently decay to a pair of lighter Higgs bosons, i.e., $gg\!\to\! H\!\to\! hh$.
Furthermore, the heavy Higgs boson is assumed to have a width significantly smaller than the detector resolution, which is approximately 1.5\% in the best case (the $\bbyy$ analysis).  The potential interference between nonresonant and resonant production is ignored.

This paper is organized as follows. For the $\bbtt$ and $\yyWW$ analyses, data and Monte Carlo~(MC) samples are described in Sec.~\ref{sec:samples} and the object reconstruction and identification are outlined in Sec.~\ref{sec:objects}. In Secs.~\ref{sec:bbyy} and~\ref{sec:bbbb}, the separately published $\bbyy$ and $\bbbb$ analyses are briefly summarized. The $\bbtt$ and $\yyWW$ analyses including event selection, background estimations and systematic uncertainties are presented in Secs.~\ref{sec:bbtt} and~\ref{sec:yyWW}, respectively. The statistical and combination procedure is described in Sec.~\ref{sec:procedure}. The results of the $\bbtt$ and $\yyWW$ analyses, as well as their combinations with the published analyses are reported in Sec.~\ref{sec:results}. The implications of the resonant search for two specific scenarios of the MSSM, hMSSM~\cite{Djouadi:2013uqa,Djouadi:2015jea} and low-tb-high~\cite{LHCtbNote}, are discussed in Sec.~\ref{sec:interpretation}. These scenarios make specific assumptions and/or choices of MSSM parameters to accommodate the observed Higgs boson. Finally, a summary is given in Sec.~\ref{sec:summary}.

\section{Data and Monte Carlo samples}
\label{sec:samples}

The data used in the searches were recorded in 2012 with the ATLAS detector at the Large Hadron Collider in  proton--proton collisions at a center-of-mass energy of 8~TeV and correspond to an integrated luminosity of 20.3 \ifb. 
The ATLAS detector is described in detail in Ref.~\cite{Aad:2008zzm}. Only data recorded when all subdetector systems were properly functional are used.

Signal and background MC samples are simulated with various event generators, each interfaced to {\sc Pythia}~v8.175~\cite{Sjostrand:2007gs} for parton showers, hadronization and underlying-event simulation. Parton distribution functions~(PDFs) CT10~\cite{Lai:2010vv} or CTEQ6L1~\cite{Pumplin:2002vw} for the proton are used depending on the generator in question. MSTW2008~\cite{Martin:2009iq} and NNPDF~\cite{Ball:2012wy} PDFs are used to evaluate systematic uncertainties. Table~\ref{tab:MCGenerators} gives a brief overview of the event generators, PDFs and cross sections used for the $\bbtt$ and $\yyWW$ analyses. All MC samples are passed through the ATLAS detector simulation program~\cite{Aad:2010ah} based on {\sc GEANT4}~\cite{Agostinelli:2002hh}. 

 \begin{table}[htb]
   \caption{List of MC generators and  parton distribution functions of the signal and background processes used by the $\bbtt$ and $\yyWW$ analyses. SM cross sections used for the normalization are also given. For the $WZ$ and $ZZ$ processes, contributions from $\gamma^*$ are included and the cross sections quoted are for $m_{Z/\gamma^*}>20$~GeV.}\vspace*{-0.3cm}
 \begin{center}
 \begin{tabular}{clccc} \hline\hline
   & Process                       & \hsc  Event generator \hsc  &  \hsd PDF set \hsd  & \hsb Cross section [pb] \hsb \\ \hline \\[-3mm]
   & \multicolumn{4}{c}{Background processes} \\ \hline
   & $V+$jets                      & {\sc Alpgen + Pythia8}      &  CTEQ6L1            & normalized to data\\ 
   & Diboson: $WW$                 & {\sc Powheg + Pythia8}      &  CT10               & 55.4 \\
   & Diboson: $WZ$                 & {\sc Powheg + Pythia8}      &  CT10               & 22.3 \\
   & Diboson: $ZZ$                 & {\sc Powheg + Pythia8}      &  CT10               & 7.3 \\[2mm]
   
   & $t\bar{t}$                    & {\sc Powheg + Pythia8}      &  CT10               & 253\\
   & Single top: $t$-channel       & {\sc AcerMC + Pythia8}      &  CTEQ6L1            & 87.8\\
   & Single top: $s$-channel       & {\sc Powheg + Pythia8}      &  CT10               & 5.6\\
   & Single top: $Wt$              & {\sc Powheg + Pythia8}      &  CT10               & 22.0\\[2mm]
   
   & $gg\!\to\! h$                 & {\sc Powheg + Pythia8}      &  CT10               & 19.2\\ 
   & $q\bar{q}'\!\to\! q\bar{q}'h$ & {\sc Powheg + Pythia8}      &  CT10               & 1.6\\
   & $q\bar{q}\!\to\! Vh$          & {\sc Pythia8}               &  CTEQ6L1            & 1.1\\
   & $q\bar{q}/gg\!\to\! t\bar{t}h$& {\sc Pythia8}               &  CTEQ6L1            & 0.13\\ \hline \\[-3mm]
   
   & \multicolumn{4}{c}{Signal processes} \\ \hline
   & Nonresonant $gg\!\to\! hh$   & {\sc MadGraph5 + Pythia8}   & CTEQ6L1             & 0.0099   \\ 
   & Resonant $gg\!\to\!H\!\to\!hh$& {\sc MadGraph5 + Pythia8}   & CTEQ6L1             & model dependent   \\
   \hline\hline
 \end{tabular}
 \end{center}
 \label{tab:MCGenerators}
 \end{table}

Signal samples for both nonresonant and resonant Higgs boson pair production are generated using the leading-order {\sc MadGraph5}~v1.5.14~\cite{Alwall:2011uj} program. The nonresonant production is modeled using the SM DiHiggs model~\cite{Plehn:1996wb,Frederix:2014hta} while the resonant  production is realized using the HeavyScalar model~\cite{HeavyScalar}, both implemented in {\sc MadGraph5}. The heavy scalar $H$ is assumed to have a narrow decay width of 10~MeV, much smaller than the experimental resolution. The SM prediction for the nonresonant $gg\!\to\! hh$ production cross section is $9.9\pm 1.3$~fb~\cite{deFlorian:2013jea} with $m_h=125.4$~GeV from the next-to-next-to-leading-order calculation in QCD.

Single SM Higgs boson production is considered as a background. The {\sc Powheg}~r{\sc 1655} generator~\cite{Nason:2004rx,Frixione:2007vw,Alioli:2010xd} is used to produce gluon fusion (ggF) and vector-boson fusion  (VBF) events. This generator calculates QCD corrections up to next-to-leading order~(NLO), including finite bottom- and top-quark mass effects~\cite{Grazzini:2013mca}. The Higgs boson transverse momentum ($\pT$) spectrum of the ggF process is matched to the calculated spectrum at next-to-next-to-leading order~(NNLO) and next-to-next-to-leading logarithm (NNLL)~\cite{deFlorian:2012mx} in QCD corrections.  Events of associated production $q\bar{q}\!\to\! Vh$ (here $V=W$ or $Z$) and $q\bar{q}/gg\!\to\! t\bar{t}h$ are produced using the {\sc Pythia8} generator~\cite{Sjostrand:2007gs}. All of these backgrounds are normalized using the state-of-the-art theoretical cross sections (see Table~\ref{tab:MCGenerators}) and their uncertainties compiled in Refs.~\cite{Dittmaier:2011ti,Dittmaier:2012vm,Heinemeyer:2013tqa}.

The {\sc Alpgen}~v2.1.4 program~\cite{Mangano:2002ea}  is used to produce the $V+$jets samples. The {\sc Powheg} generator is used to simulate top quark pair production ($t\bar{t}$) as well as the $s$-channel and $Wt$ processes of single top production; the $t$-channel process of single top production is simulated using the {\sc AcerMC}~v38 program~\cite{Kersevan:2004yg}. The $t\bar{t}$ cross section has been calculated up to NNLO and NNLL in QCD corrections~\cite{Czakon:2013goa}. Cross sections for the three single-top processes have been calculated at (approximate) NNLO accuracy~\cite{Kidonakis:2011wy,Kidonakis:2010tc,Kidonakis:2010ux}. The {\sc Powheg} generator is  used to simulate diboson backgrounds ($WW$, $WZ$ and $ZZ$). The diboson production cross sections are calculated at NLO in QCD corrections using the MCFM program~\cite{Campbell:2010ff,Campbell:2011bn}.

\section{Object identification}
\label{sec:objects}

In this section,  object reconstruction and identification for the $\bbtt$ and $\yyWW$ analyses are discussed. The $\bbtt$ and $\yyWW$ analyses are developed following the $h\!\to\!\tau\tau$~\cite{Aad:2015vsa} and $h\!\to\!\gamma\gamma$~\cite{atlas:2014hgg} studies of single Higgs bosons, respectively, and use much of their methodology.

Electrons are reconstructed from energy clusters in the electromagnetic calorimeter matched to tracks in the inner tracker. The calorimeter shower profiles of electron candidates must be consistent with those expected from electromagnetic interactions. Electron candidates are identified using tight and medium criteria~\cite{ATLAS:2014wga} for the $\bbtt$ and $\yyWW$ analyses, respectively. The selected candidates  are required to have transverse momenta\footnote{ATLAS uses a right-hand coordinate system with the interaction point as its origin and the beam line as its $z$-axis. The $x$-axis points to the center of the LHC ring and $y$-axis points upwards. The pseudorapidity $\eta$ is defined as $\eta = -\ln \tan(\theta/2)$, where $\theta$ is the polar angle measured with respect to the $z$-axis. The transverse momentum $\pT$ is calculated from the momentum $p$: $\pT=p\sin\theta$.} $\pT>15$~GeV and be within the detector fiducial volume of $|\eta|<2.47$ excluding $1.37<|\eta|<1.52$, the transition region between the barrel and endcap calorimeters. Muons are identified by matching tracks or segments reconstructed in the muon spectrometer with tracks reconstructed in the inner tracker. They are required to have $\pT>10$~GeV and $|\eta|<2.5$. Both the electrons and muons must satisfy calorimeter and track isolation requirements. The calorimeter isolation requires that the energy deposited in the calorimeter in a cone of size $\Delta R\equiv\sqrt{(\Delta\eta)^2+(\Delta\phi)^2}=0.2$ around the lepton (electron or muon), excluding the energy deposited by the lepton itself, is less than 6\% (20\%) of the $\pT$ of the lepton for the $\bbtt$ ($\yyWW$) analysis. The track isolation is defined similarly: the scalar $\pT$ sum of additional tracks originating from the primary vertex with $\pT>1$~GeV in a cone of size $\Delta R = 0.4$ around the lepton is required to be less than 6\% (15\%) of the $\pT$ of the lepton track for the $\bbtt$ ($\yyWW$) analysis.

Photons are reconstructed from energy clusters in the electromagnetic calorimeter with their shower profiles consistent with electromagnetic showers. A significant fraction of photons convert into $e^+e^-$ pairs inside the inner tracker. Thus photon candidates are divided into unconverted and converted categories. Clusters without matching tracks are considered as unconverted photons, while clusters matched to tracks consistent with conversions are considered as converted photons. Photon candidates must fulfill the tight identification criteria~\cite{ATLAS:2012ana} and  are required to have $\pT>25$~GeV and $|\eta|<2.37$ (excluding the transition region $1.37<|\eta|<1.52$) and must satisfy both calorimeter and track isolation. The calorimeter isolation requires the additional energy in a cone of $\Delta R=0.4$ around the photon candidate to be less than 6~GeV while the track isolation requires the scalar $\pT$ sum of additional tracks in a cone of $\Delta R=0.2$ around the photon to be less than 2.6~GeV.

Jets are reconstructed using the anti-$k_t$ algorithm~\cite{Cacciari:2008gp} with a radius parameter of $R=0.4$. Their energies are corrected for the average contributions from pileup interactions. Jets are required to have $\pT>30$~GeV and $|\eta|<4.5$. For the $\yyWW$ analysis, a lower $\pT$ requirement of 25~GeV is applied for jets in the central region of $|\eta|<2.4$. 
To suppress contributions from pileup interactions, jets with
$\pT<50$~GeV and within the acceptance of the inner tracker
($|\eta|<2.4$) must have over 50\% of the scalar sum of the $\pT$ of
their associated tracks contributed by those originating from the
primary vertex.
Jets containing $b$-hadrons are identified using a multivariate algorithm ($b$-tagging)~\cite{ATLAS:2014pla}.  The algorithm combines information such as the explicit reconstruction of the secondary decay vertices and track impact-parameter significances. The operating point chosen for both $\bbtt$ and $\yyWW$ analyses has an efficiency of 80\% for the $b$-quark jets in $t\bar{t}$ events.

Hadronically decaying $\tau$ candidates ($\tauhad$) are reconstructed using clusters in the electromagnetic and hadronic calorimeters~\cite{Aad:2014rga}. The tau candidates are required to have $\pT>20$~GeV and $|\eta|<2.5$. The number of tracks with $\pT>1$~GeV associated with the candidates must be one or three and the total charge determined from these tracks must be $\pm 1$. The tau identification uses calorimeter cluster as well as  tracking-based variables, combined using a boosted-decision-tree method~\cite{Aad:2014rga}. Three working points, labeled loose, medium and tight~\cite{Aad:2014rga}, corresponding to different identification efficiencies are used. Dedicated algorithms that suppress electrons and muons misidentified as $\tauhad$ candidates are applied as well.

The missing transverse momentum (with magnitude $\met$) is the negative of the vector sum of the transverse momenta of all photon, electron, muon, $\tauhad$, and jet candidates, as well as the $\pT$ of all calorimeter clusters not associated with these reconstructed objects, called the soft-term contribution~\cite{Aad:2012re}. 
The $\bbtt$ analysis uses the version of the $\met$ calculation in the $h\!\to\!\tau\tau$ analysis~\cite{Aad:2015vsa}. In this calculation, the soft-term contribution  is scaled by  a vertex fraction, defined as the ratio of the summed scalar $\pT$ of all tracks from the primary vertex not matched with the reconstructed objects to the summed scalar $\pT$ of all tracks in the event. 
The $\yyWW$ analysis, on the other hand, uses the $\met$-significance employed by the $h\!\to\!\gamma\gamma$ study~\cite{atlas:2014hgg}. It is defined as the ratio of the measured $\met$ to its expected resolution estimated using the square root of the scalar sum of the transverse energies of all objects contributing to the $\met$ calculation.

\section{Summary of $\bbyy$}
\label{sec:bbyy}

The $\bbyy$ analysis, published in Ref.~\cite{Aad:2014yja}, largely follows the ATLAS analysis of the Higgs boson discovery in the $\hyy$ decay channel~\cite{atlas:2012obs,atlas:2014hgg}. The search is performed in the $\sqrt{s}=8$~TeV dataset corresponding to an integrated luminosity of 20.3~\ifb. The data were recorded with  diphoton triggers that are nearly 100\% efficient for events satisfying the photon requirements. Events must contain two isolated photons. The $\pT$ for the leading (subleading) photon must be larger than 35\% (25\%) of the diphoton invariant mass $\myy$, which itself must be in the range of $105 < \myy < 160$~GeV. Events must also contain two energetic $b$-tagged jets; the leading (subleading) jet must have $\pt > 55$ (35)~GeV, and the dijet mass must fall within a window $95 < \mbb < 135$~GeV, as expected from the $\hbb$ decay. A multivariate $b$-tagging algorithm~\cite{ATLAS:2014pla} that is 70\% efficient for the $b$-quark jets in \ttbar\ events is applied.

Backgrounds for both the resonant and nonresonant analyses are divided into two categories: events containing a single real Higgs boson (with $h \rightarrow \gamma \gamma$), and the continuum background of events not containing a Higgs boson. The former are evaluated purely from simulation, and are small compared to the continuum background, which is evaluated from data in the diphoton mass sidebands (the $\myy$ range of 105--160~GeV excluding the region of $m_h\pm 5$~GeV). In the nonresonant analysis, an unbinned signal-plus-background fit is performed on the observed $\myy$ distribution, with the background from single Higgs bosons constrained to the expectation from the SM. The continuum background is modeled with an exponential function; the shape of the exponential function is taken from data containing a diphoton and dijet pair where fewer than two jets are $b$-tagged. 

The resonant search proceeds in a similar manner, although it is ultimately a counting experiment, with an additional requirement on the four-object invariant mass $\myybb$, calculated with the $\bb$ mass constrained to $m_h$. This requirement on $\myybb$ varies with the resonance mass hypothesis under evaluation, and is defined as the smallest window containing 95\% of the signal events based on MC simulation. As in the nonresonant search, the number of background events with real Higgs bosons is estimated from simulation. The continuum background in the $\myy$ signal window is extrapolated from the diphoton mass sidebands.
A resonance with mass between 260~GeV and 500~GeV is considered in the search.

The small number of events (nine) in the diphoton mass sideband leads to large statistical uncertainties (33\%) on the dominant continuum background, so that most systematic uncertainties have a small effect on the final result. 
For the resonant search, however, systematic uncertainties with comparable effect remain. Uncertainties of 0--30\% (depending on the resonance mass hypothesis under consideration) are assigned due to the modeling of the $\myybb$ shape using the data with less than two $b$-tagged jets. Additional uncertainties of 16--30\% arise from the choice of functional form used to parameterize the shape of $\myybb$.

In the nonresonant analysis, extrapolating the sidebands into the diphoton mass window of the signal selection leads to a prediction of 1.3 continuum background events. An additional contribution of 0.2 events is predicted from single Higgs boson production. A total of five events are observed, representing an excess of 2.4~standard deviations~($\sigma$). 
A 95\% confidence level~(CL) upper limit of 2.2~(1.0)~pb is observed (expected) for $\sigma(gg\!\to\! hh)$, the cross section of nonresonant Higgs boson pair production. For the resonant searches, the observed (expected) upper limits on $\sigma(gg\!\to\! H)\times {\rm BR}(H\!\to\! hh)$ are 2.3~(1.7)~pb at $m_H=260$~GeV  and 0.7~(0.7)~pb at $m_H=500$~GeV.

\section{Summary of $\bbbb$}
\label{sec:bbbb}
The $\bbbb$ analysis~\cite{Aad:2015uka} benefits from the large branching ratio of $h\!\to\!\bb$. The analysis employs resolved as well as boosted Higgs boson reconstruction methods. The resolved method attempts to reconstruct and identify separate $b$-quark jets from the $\hbb$ decay, while the boosted method uses a jet substructure technique to identify the $\hbb$ decay reconstructed as a single jet. The latter is expected if the Higgs boson $h$ has a high momentum. The boosted method is particularly suited to the search for a resonance with mass above approximately 1000~GeV decaying to a pair of SM Higgs bosons.
For the combinations presented in this paper, resonances below this mass are considered and the resolved method is used as it is more sensitive.

The analysis with the resolved method searches for two back-to-back and high-momentum $\bb$ systems with their masses consistent with $m_h$ in a dataset at $\sqrt{s}=8$~TeV corresponding to an integrated luminosity of 19.5~\ifb\ for the triggers used. The data were recorded with a combination of multijet triggers using information including the $b$-quark jet tagging. The trigger is $>\!99.5\%$ efficient for signal events satisfying the offline selection. Candidate events are required to have at least four $b$-tagged jets, each with $\pT>40$~GeV. As in the $\bbyy$ analysis, a multivariate $b$-tagging algorithm~\cite{ATLAS:2014pla} with an estimated efficiency of 70\% is used to tag jets containing $b$-hadrons. The four highest-$\pT$ $b$-tagged jets are then used to form two dijet systems, requiring the angular separation $\Delta R$ in $(\eta, \phi)$ space between the two jets in each of the two dijet systems to be smaller than 1.5. The transverse momenta of the leading and subleading dijet systems must be greater than 200~GeV and 150~GeV, respectively. These selection criteria, driven partly by the corresponding jet trigger thresholds and partly by the necessity to suppress the backgrounds, lead to significant loss of signal acceptance for lower resonance masses. The resonant search only considers masses above 500~GeV. The leading ($m_{12}$) and subleading ($m_{34}$) dijet invariant mass values are required to be consistent with those expected for the $\bbbb$ decay, satisfying the requirement:
$$\sqrt{\left(\frac{m_{12}-m_{12}^0}{\sigma_{12}}\right)^2+\left(\frac{m_{34}-m_{34}^0}{\sigma_{34}}\right)^2}<1.6.$$
Here $m_{12}^0$ (124 GeV) and $m_{34}^0$ (115 GeV) are the expected peak values from simulation for the leading and subleading dijet pair, respectively, and $\sigma_{12}$ and $\sigma_{34}$ are the dijet mass resolutions, estimated from the simulation to be 10\% of the dijet mass values. More details about the selection can be found in Ref.~\cite{Aad:2015uka}.

After the full selection, more than 90\% of the total background in the signal region is estimated to be multijet events, while the rest is mostly $t\bar{t}$ events. The $Z$+jets background constitutes less than 1\% of the total background and is modeled using MC simulation.  The multijet background is modeled using a fully data-driven approach in an independent control sample passing the same selection as the signal except that only one of the two selected dijets is $b$-tagged. This control sample is corrected for the kinematic differences arising from the additional $b$-tagging requirements in the signal sample. The $t\bar{t}$ contribution is taken from MC simulations normalized to data in dedicated control samples.

The dominant sources of systematic uncertainty in the analysis are the $b$-tagging calibration and the modeling of the multijet background. The degradation in the analysis sensitivity from these uncertainties is below 10\%. Other sources
of systematic uncertainty include the $t\bar{t}$ modeling, and the jet energy scale and resolution, which are all at the percent level. 
A total of 87 events are observed in the data, in good agreement with the SM expectation of $87.0\pm 5.6$ events. Good agreement is also observed in the four-jet invariant mass distribution, thus there is no evidence of  Higgs boson pair production. For the  nonresonant search, both the observed and expected 95\% CL upper limit on the cross section $\sigma(pp\!\to\! \bbbb)$ is 202~fb.  
For the resonant search, the invariant mass of the four jets is used as the final discriminant from which the upper limit on the potential signal cross section is extracted. The resulting observed (expected) 95\% CL upper limit on $\sigma(pp\!\to\! H\!\to\! \bbbb)$ ranges from 52~(56)~fb, at $m_H=500$~GeV, to 3.6~(5.8)~fb, at $m_H=1000$~GeV.

\section{$\bbtt$}
\label{sec:bbtt}

This section describes the search for Higgs boson pair production in the $\bbtt$ decay channel, where only the final state where one tau lepton decays hadronically and the other decays leptonically,  $bb\ell\tau_{\rm had}$, is used. The data were recorded with triggers requiring at least one lepton with $p_{\rm T}>24$~GeV. These triggers are nearly 100\% efficient for events passing the final selection. Candidate $\bblt$ events are selected by requiring exactly one lepton  with $\pT>26$~GeV, one hadronically decaying tau lepton of the opposite charge with $\pT>20$~GeV meeting the medium criteria~\cite{Aad:2014rga}, and two or more jets with $\pT>30$~GeV. In addition, between one and three of the selected jets must be $b$-tagged using the multivariate $b$-tagger. The upper bound on the number of $b$-tagged jets is designed to make this analysis statistically independent of the $\bbbb$ analysis summarized in Sec.~\ref{sec:bbbb}. 
These criteria are collectively referred to as the ``preselection''. 

The backgrounds from $W$+jets, $Z\!\to\!\tau\tau$, diboson ($WW,\,WZ$ and $ZZ$) and top quark (both $t\bar{t}$ and single top quark) production dominate the surviving sample and their contributions are derived from a mixture of data-driven methods and simulation. The contribution from events with a jet misidentified as a $\tauhad$, referred to as the fake $\tauhad$ background, are estimated using data with a ``fake-factor'' method. The method estimates contributions from $W$+jets, multijet, $Z$+jets and top quark events that pass the event selection due to misidentified $\tauhad$ candidates. The fake factor is defined as the ratio of the number of $\tauhad$ candidates identified as medium, to the number satisfying the loose, but not the medium, criteria~\cite{Aad:2014rga}. The $\pT$-dependent fake factors are measured in data control samples separately for the $\tauhad$ candidates with one or three tracks and for $W$+jets, multijet, $Z$+jets and top quark contributions. The $W$+jets, multijet, $Z$+jets and top quark control samples are selected by reversing the $m_T$ cut (see below), relaxing the lepton isolation requirement, reversing the dilepton veto or by requiring $b$-tagged jets, respectively. The fake factors determined from these control samples are consistent within their statistical uncertainties. They are then applied to the signal control sample, i.e.,  events passing the selection, except that the $\tauhad$  candidate is required to satisfy the loose, but not the medium, $\tauhad$ identification, to estimate the fake $\tauhad$ background. The composition of the sample is determined from a mixture of data-driven methods and MC simulations and it is found that the sample is dominated by the $W$+jets and multijet events. Details of the method can be found in Ref.~\cite{Aad:2014rga}. The method is validated using the same-sign $\ell\tauhad$ events that are otherwise selected in the same way as the signal candidates.

The $Z\rightarrow\tau\tau$ background is modeled using selected $Z\rightarrow\mu\mu$ events in data through embedding~\cite{Aad:2015kxa}, where the muon tracks and associated energy depositions in the calorimeters are replaced by the corresponding simulated signatures of the final-state particles of tau decays. In this approach, the kinematics of the produced boson, the hadronic activity of the event (jets and underlying event) as well as pileup are taken from data~\cite{Aad:2015vsa}. Other processes passing the $Z\!\to\!\mu\mu$ selection, primarily from top quark production, are subtracted from the embedded data sample using simulation. Their contributions are approximately 2\% for events with one $b$-tagged jet and 25\% for events with two or more $b$-tagged jets. The $Z\to\tau\tau$ background derived is found to be in a good agreement with that obtained from the MC simulation. 

The remaining backgrounds, mostly $t\bar{t}$ and diboson events with genuine $\ell\tauhad$ in their decay final states, are estimated using simulation. The small contributions from single SM Higgs boson production and from $Z(\!\to\! ee/\mu\mu)+{\rm jets}$ events (in which one of the electrons or muons is misidentified as $\tauhad$) are also estimated from simulation. The production rates of these processes are normalized to the theoretical cross sections discussed in Sec.~\ref{sec:samples}. For the simulation of the $t\bar{t}$ process, the top quark $\pT$ distribution is corrected for the observed difference between data and simulation~\cite{Aad:2014zka}. The background from misidentified leptons is found to be negligible.

\begin{figure}[!h!tpb]
  \centering
  \begin{tabular}{cc}
\subfloat[]{\includegraphics[width=0.48\textwidth]{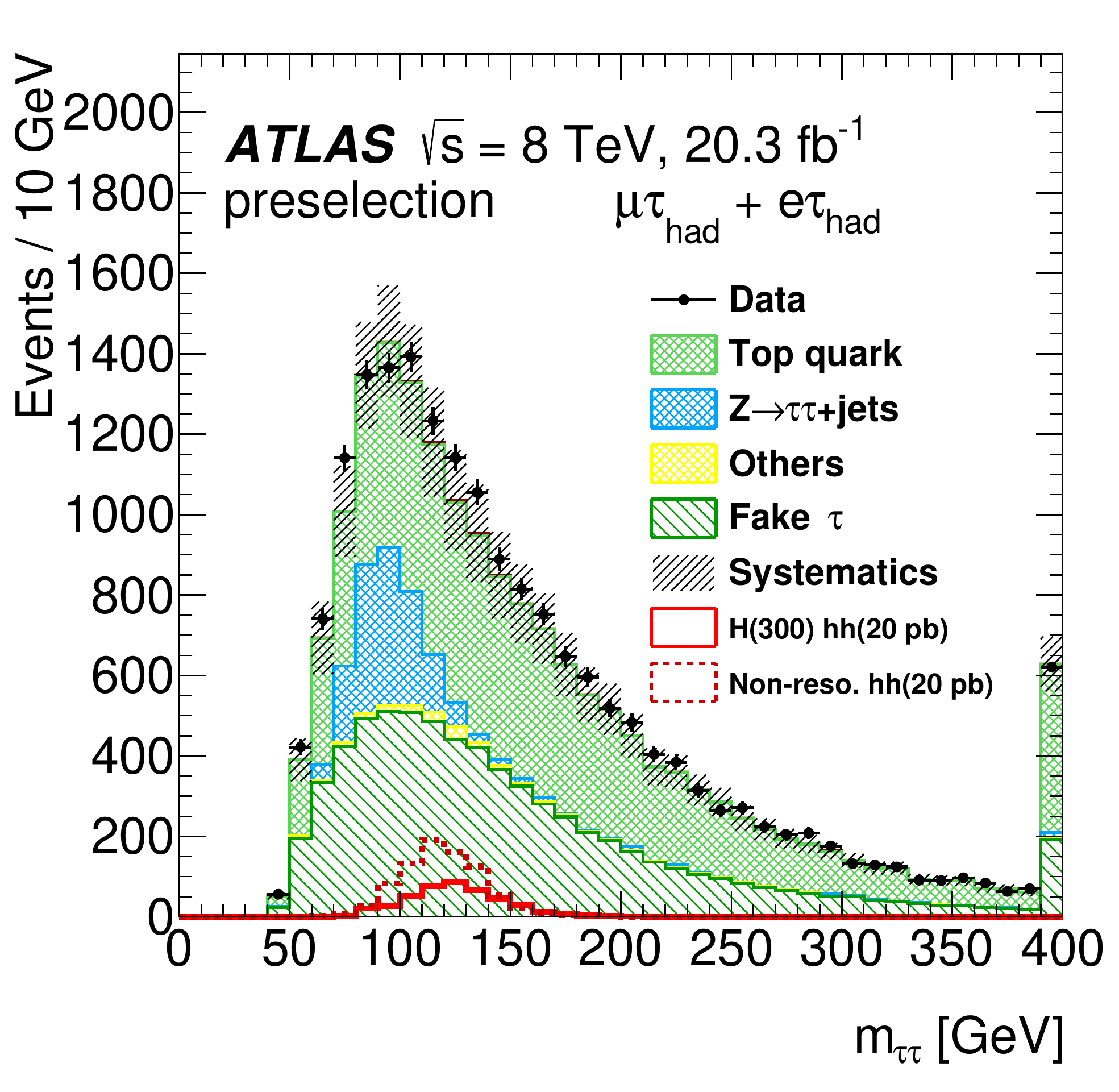}\label{fig:bbtt-pre_mmc}} &
\subfloat[]{\includegraphics[width=0.48\textwidth]{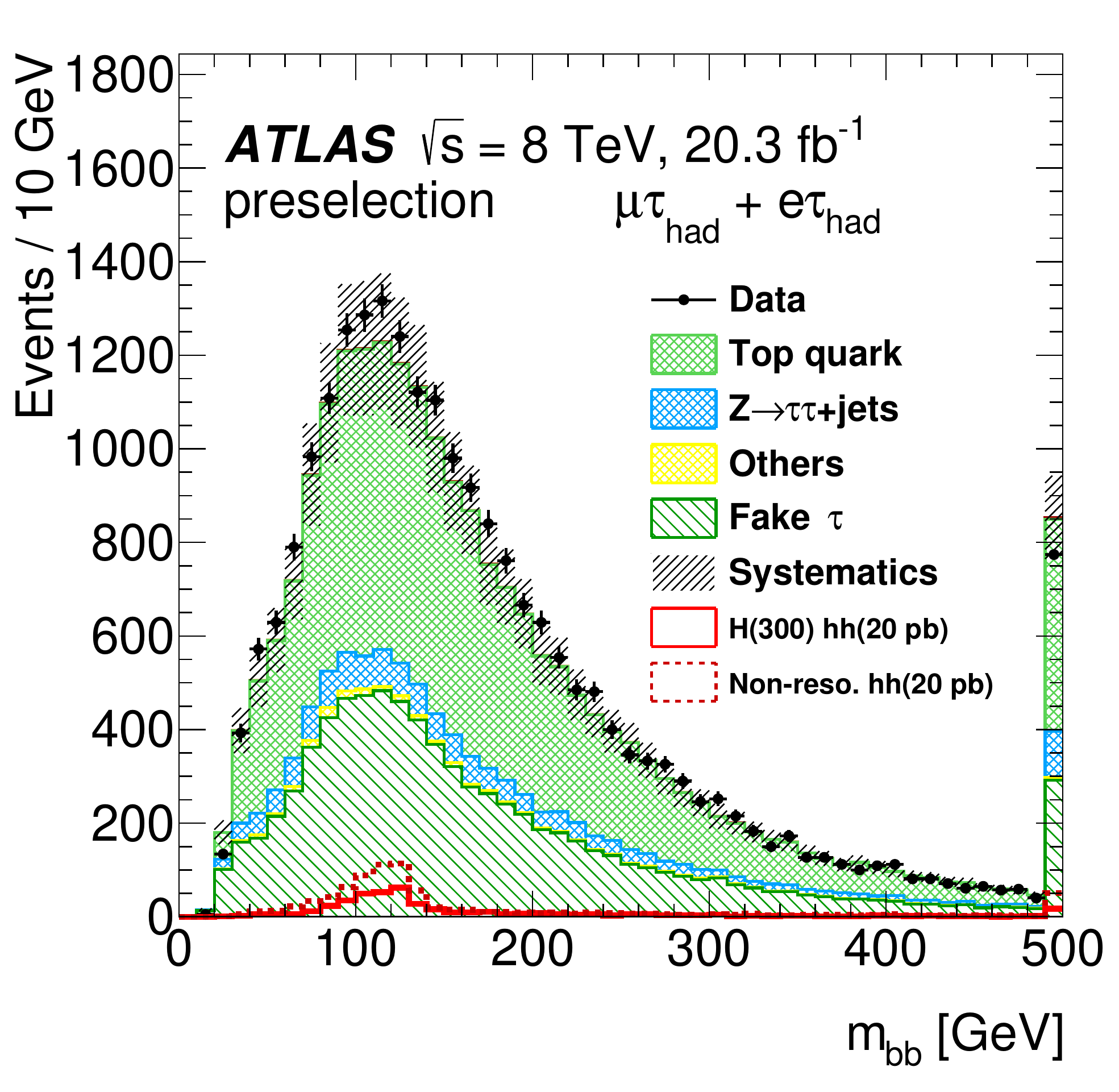}\label{fig:bbtt-pre_mjj}} \\
\multicolumn{2}{c}{\subfloat[]{\includegraphics[width=0.48\textwidth]{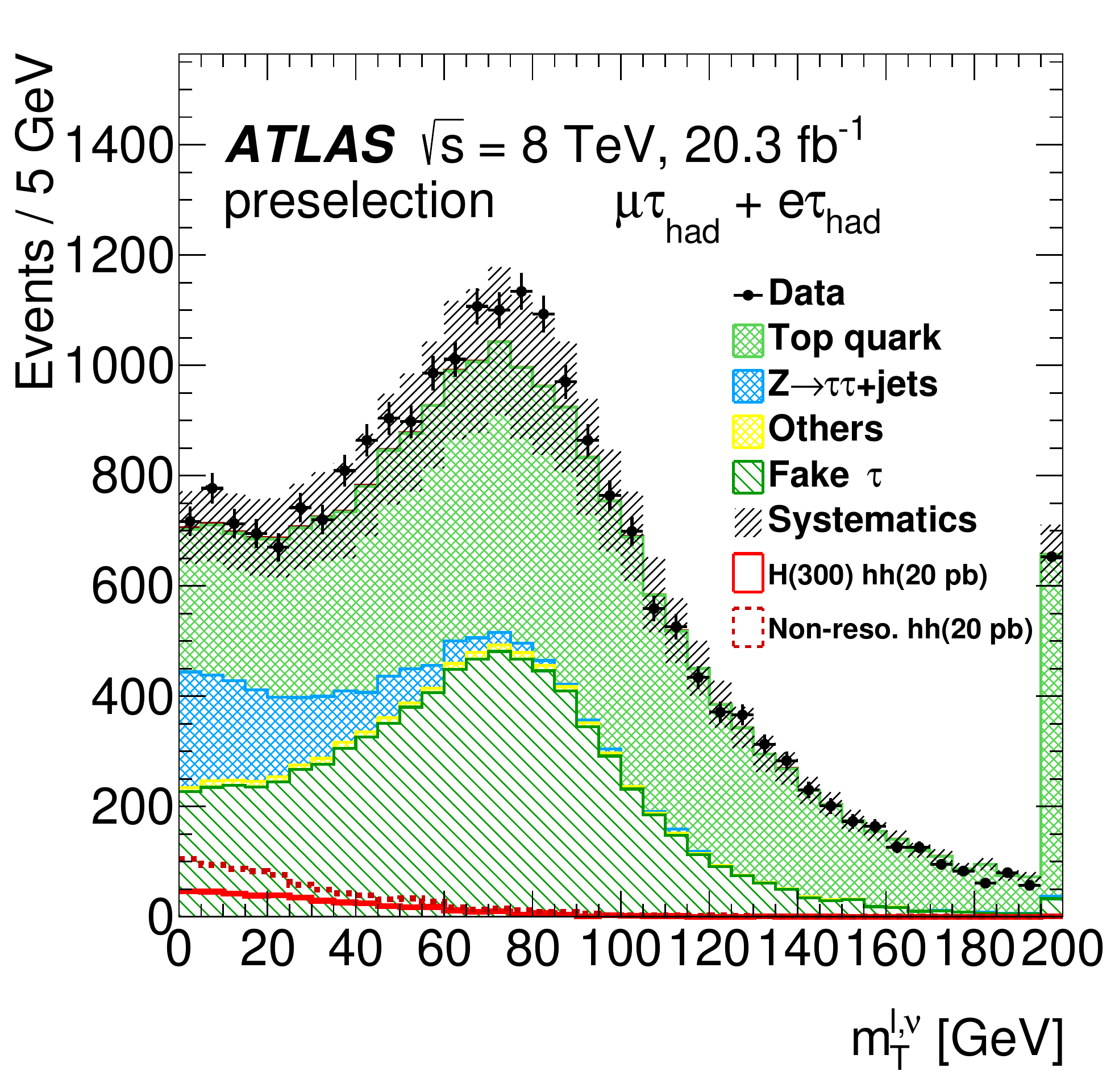}\label{fig:bbtt-pre_mt}}} \\
  \end{tabular}
  \caption{Kinematic distributions of the $\bbtt$ analysis after the preselection (see text) comparing data with the expected background contributions: (a) ditau mass $\mtt$ reconstructed using the MMC method, (b) dijet mass $\mbb$ and (c) the transverse mass $\mTlv$ of the lepton and the $\met$ system. 
The top quark background includes contributions from both $t\bar{t}$ and the single top-quark production. 
The background category labeled ``Others'' comprises diboson and $Z\!\to\! ee/\mu\mu$ contributions. 
Contributions from single SM Higgs boson production are included in the background estimates, but are too small to be visible on these distributions. As illustrations, expected signal distributions for a Higgs boson pair production cross section of 20~pb are overlaid for both nonresonant and resonant Higgs boson pair production. A mass of $m_H=300$~GeV is assumed for the resonant production. The last bin in all distributions represents overflows. The gray hatched bands represent the uncertainties on the total background contributions. These uncertainties are largely correlated from bin to bin.
}
\label{fig:bbtt-pre}
\end{figure}

Figures~\ref{fig:bbtt-pre}(a) and~\ref{fig:bbtt-pre}(b) compare the observed ditau ($\mtt$) and dijet ($\mbb$) mass distributions with those expected from background events after the preselection discussed above. The sample is dominated by contributions from top quark production, fake $\tauhad$, and $Z\!\to\!\tau\tau$ backgrounds. Also shown in the figures are the expected signal distributions for a Higgs boson pair production cross section of 20~pb as an illustration. 
The yield of the nonresonant production is significantly higher than that of the resonant production for the same cross section, largely due to the harder $\pT$ spectrum of the Higgs boson $h$ of the nonresonant production.
The ditau invariant mass is reconstructed from the electron or muon, $\tauhad$, and $\met$ using a method known as the missing mass calculator (MMC)~\cite{Elagin:2010aw}. The MMC solves an underconstrained system of equations with solutions weighted by $\met$ resolution and the tau-lepton decay topologies. It returns the most probable value of the ditau mass, assuming that the observed lepton, $\tauhad$ and $\met$ stem from a $\tau\tau$ resonance. 
The dijet mass is calculated from the two leading $b$-tagged jets, or using also the highest-$\pT$ untagged jet if only one jet is $b$-tagged.

Additional topological requirements are applied to reduce the background. As shown in Fig.~\ref{fig:bbtt-pre}(c), the signal events tend to have small values of the transverse mass $\mTlv$ calculated from the lepton and $\met$ system. Consequently, a requirement of $\mTlv<60$~GeV is applied, which  reduces the background significantly with only a small loss of the signal efficiency. In addition, the angular separation in the transverse plane between the $\met$  and $\tauhad$ is required to be larger than one radian to reduce the fake $\tauhad$ background. 

Background events from  $t\bar{t}\!\to\! WW\bb\!\to\!\ell\nu\tau\nu \bb$ decay have an identical visible final state to the signal if the tau lepton decays hadronically. For signal $h\!\to\!\tau\tau\!\to\!\ell\tauhad$ events, however, the $\pT$ of the lepton tends to be softer than that of the $\tauhad$ due to the presence of more neutrinos in the $\tau\!\to\!\ell$ decays. Thus the $\pT$ of the electron or muon is required to satisfy $\pT(\ell)< \pT(\tauhad)+20\,{\rm GeV}$. The $t\bar{t}$ events of the $t\bar{t}\!\to\! WW\,\bb \!\to\! \ell\nu\,qq'\,\bb$ final state with a misidentified $\tauhad$ candidate remain a large background. To reduce its contribution, a $W$ boson candidate is reconstructed from the $\tauhad$ candidate and its closest untagged jet and its mass $m_{\tau j}$ is calculated. The $W$ candidate is then paired with a $b$-tagged jet to form the top quark candidate with a reconstructed mass $m_{\tau j b}$. The pairing is chosen to minimize the mass sum $m_{\ell b}+m_{\tau b}$ for events with two or more $b$-tagged jets. If only one jet is $b$-tagged, one of the $b$-jets in the mass sum is replaced by the highest-$\pT$ untagged jet.
An elliptical requirement in the form of a $\chi^2$ in the $(m_{\tau j},\, m_{\tau jb})$ plane:
$$\left(\frac{ \Delta m_W\cos\theta - \Delta m_t\sin\theta}{28\,\GeV}\right)^2+\left(\frac{\Delta m_W\sin\theta+ \Delta m_t\cos\theta}{18\,\GeV}\right)^2 >1$$
is applied to reject events with  $(m_{\tau j},\,m_{\tau j b})$ consistent with the hypothesis $(m_W,\, m_t)$, the masses of the $W$ boson and the top quark. Here $\Delta m_W = m_{\tau j}-m_W,\, \Delta m_t = m_{\tau j b}-m_t$ and $\theta$ is a rotation angle given by $\tan\theta=m_t/m_W$ to take into account the average correlation between $m_{\tau j}$ and $m_{\tau j b}$.

Finally, the remaining events must have $90<\mbb<160$~GeV, consistent with the expectation for the $\hbb$ decay. For the nonresonant search, $\mtt$ is used as the final discriminant to extract the signal, and its distribution is shown in Fig.~\ref{fig:bbtt-SR}(a). The selection efficiency for the $gg\!\to\! hh\!\to\! \bb\tau\tau$ signal is 0.57\%. For the resonant search, the MMC mass is required to be in the range of $100<\mtt<150$~GeV. The mass resolutions of $\mbb$ and $\mtt$ are comparable for the signal, but the $\mbb$ distribution has a longer tail. The resonance mass $\mbbtt$ reconstructed from the dijet and ditau system is used as the discriminant. To improve the mass resolution of the heavy resonances, scale factors of $m_{h}$/$\mbb$ and $m_{h}$/$\mtt$ are applied respectively to the four-momenta of the $\bb$ and $\tau\tau$ systems, where $m_{h}$ is set to the value of 125 GeV used in the simulation. The resolution of $m_{bb\tau\tau}$ is found from simulation studies to vary from 2.4\% at $m_H=260$~GeV to 4.8\% at 1000~GeV. The improvement in the resolution from the rescaling is largest at low mass and varies from approximately a factor of three at 260~GeV to about 30\% at 1000~GeV. 
The reconstructed $\mbbtt$ distribution for events passing all the selections is shown in Fig.~\ref{fig:bbtt-SR}(b). The efficiency for the $gg\!\to\! H\to\! hh\!\to\!\bb\tau\tau$ signal varies from 0.20\% at 260~GeV to 1.5\% at 1000~GeV.
These efficiencies include branching ratios of the tau lepton decays, but not those of heavy or light Higgs bosons.

To take advantage of different signal-to-background ratios in different kinematic regions, the selected events are divided into four categories based on the ditau transverse momentum $\pT^{\tau\tau}$ (less than or greater than 100~GeV) and the number of $b$-tagged jets ($n_b=1$ or $\ge 2$) for both the nonresonant and resonant searches. The numbers of events expected from background processes and observed in the data passing the resonant $\bbtt$ selection are summarized in Table~\ref{tab:bbtt-yield:reso} for each of the four categories. The expected number of events from the production of a Higgs boson with $m_H=300$~GeV and a cross section of $\sigma(gg\!\to\! H)\times{\rm BR}(H\!\to\! hh)=1$~pb for each category is also shown for comparison.

\begin{table}[!htbp]
\caption{
    The numbers of events predicted from background processes and observed in the data passing the final selection of 
    the resonant search for the four categories. 
    The top quark background includes contributions from both $t\bar{t}$ and the single top-quark production. 
    The ``others'' background comprises diboson and $Z\!\to\! ee/\mu\mu$ contributions. 
    The numbers of events expected from the production of a $m_H=300$~GeV Higgs boson with a cross section 
    of $\sigma(gg\!\to\! H)\times {\rm BR}(H\!\to\! hh)=1$~pb are also shown as illustrations. 
    The uncertainties shown are the total uncertainties, combining statistical and systematic components. 
\label{tab:bbtt-yield:reso}
    }
  \begin{center}
    \begin{tabular}{ l | r@{ }r@{ }l  | r@{ }r@{ }l | r@{ }r@{ }l | r@{ }r@{ }l }\hline \hline 
                    & \multicolumn{6}{c|}{$n_b=1$}   & \multicolumn{6}{c}{$n_b\ge 2$} \\ \cline{2-7}\cline{8-13} \\[-4mm]  
    Process         & \multicolumn{3}{c|}{$\pTtt<100$ GeV } & \multicolumn{3}{c|}{ $\pTtt>100$ GeV } & \multicolumn{3}{c|}{ $\pTtt<100$ GeV } & \multicolumn{3}{c}{ $\pTtt>100$~GeV} \\ \hline

      SM Higgs        &  0.5 & $\pm$ & 0.1  &  0.8 & $\pm$ & 0.1 &  0.1 & $\pm$ & 0.1 &  0.2 & $\pm$ & 0.1 \\ 
      Top quark       & 30.3 & $\pm$ & 3.6  & 19.6 & $\pm$ & 2.5 & 30.9 & $\pm$ & 3.0 & 23.6 & $\pm$ & 2.5 \\       
      $Z\!\to\!\tau\tau$  & 38.1 & $\pm$ & 4.4  & 20.2 & $\pm$ & 3.7 &  6.8 & $\pm$ & 1.8 &  2.6 & $\pm$ & 1.0 \\       
      Fake $\tauhad$     & 37.0 & $\pm$ & 4.4  & 12.1 & $\pm$ & 1.7 & 13.7 & $\pm$ & 1.9 &  5.4 & $\pm$ & 1.0 \\ 
      Others          & 3.2  & $\pm$ & 3.7  &  0.5 & $\pm$ & 1.5 &  0.7 & $\pm$ & 1.6 &  0.2 & $\pm$ & 0.7 \\ \hline      
      Total background  &109.1 & $\pm$ & 8.6 & 53.1 & $\pm$ & 6.0 & 52.2 & $\pm$ & 8.2 & 32.1 & $\pm$ & 5.4 \\ \hline
      \multicolumn{1}{l}{Data}            & \multicolumn{3}{c}{92}  &   \multicolumn{3}{c}{46}    & \multicolumn{3}{c}{35}    & \multicolumn{3}{c}{35} \\ 
      \multicolumn{13}{c}{} \\[-3mm] \hline

      Signal \@ $m_H=300$ GeV  & 0.8 & $\pm$ & 0.2 & 0.4 & $\pm$ & 0.2 & 1.5 & $\pm$ & 0.3 & 0.9 & $\pm$ & 0.2\\
      \hline\hline\end{tabular} 
  \end{center}
\end{table}

\begin{figure}[!htpb]
\begin{center}
\subfloat[]{\includegraphics[width=0.48\textwidth]{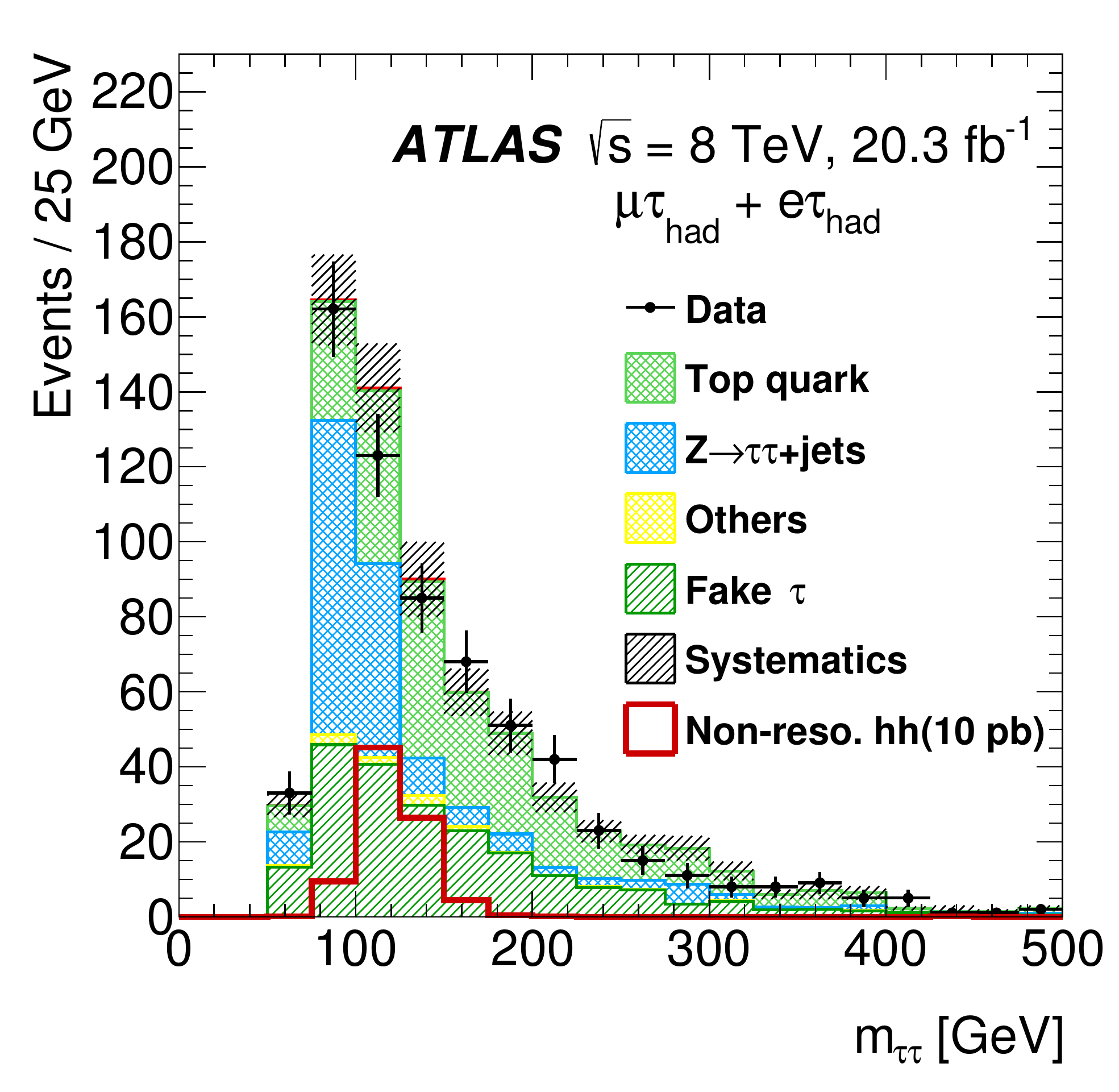}\label{fig:bbtt-SR_mmc}}\hspace*{0.5cm}
\subfloat[]{\includegraphics[width=0.48\textwidth]{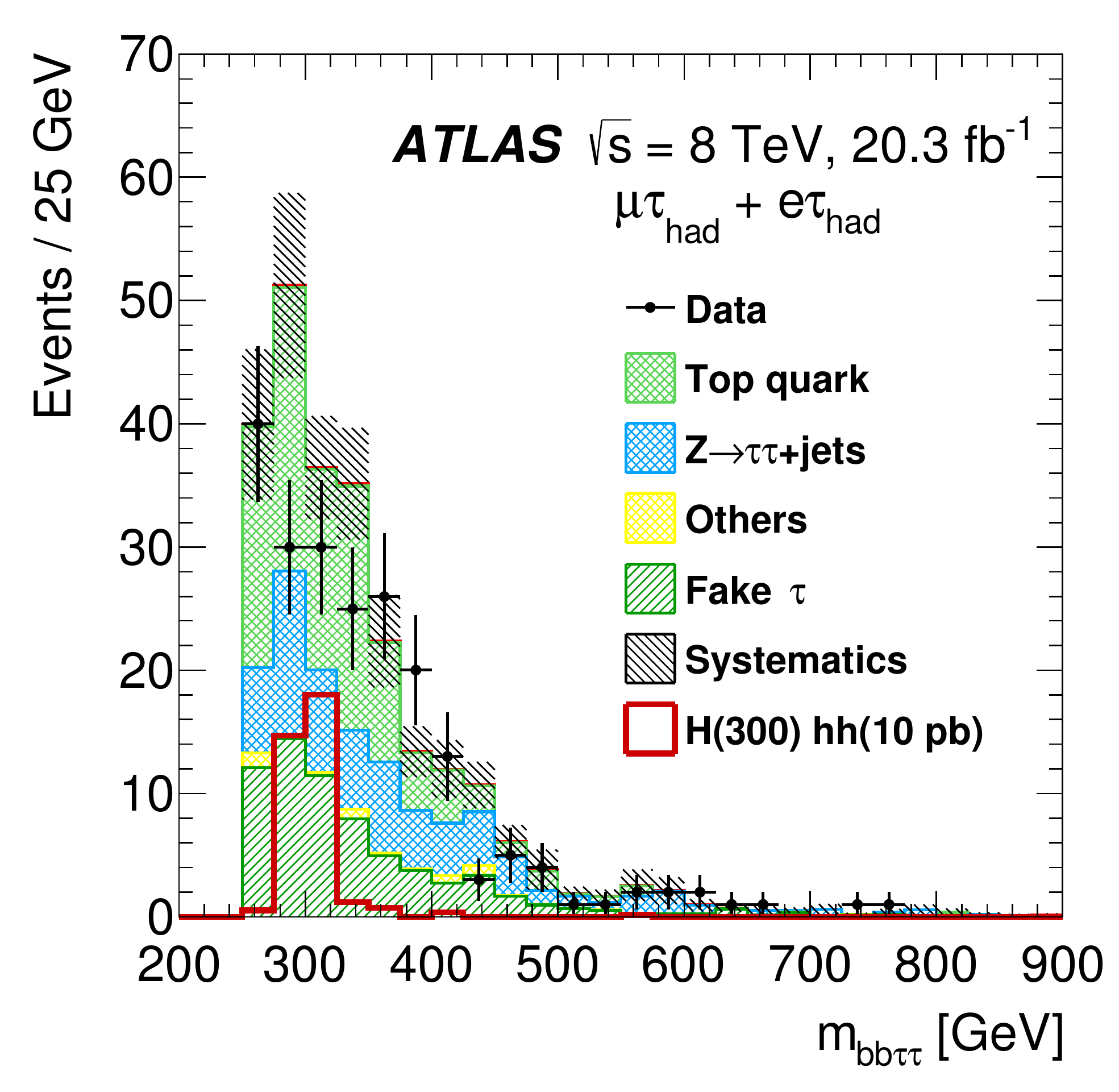}\label{fig:bbtt-SR_mbbtt}}
\end{center}
\caption{Distributions of the final discriminants used to extract the signal:  (a) $\mtt$ for the nonresonant search and (b) $\mbbtt$ for the resonant search. 
The top quark background includes contributions from both $t\bar{t}$ and the single top-quark production. 
The background category labeled ``Others'' comprises diboson and $Z\!\to\! ee/\mu\mu$ contributions. 
Contributions from single SM Higgs boson production are included in the background estimates, but are too small to be visible on these distributions. As illustrations,  the expected signal distributions assume a cross section of 10~pb for Higgs boson pair production for both the nonresonant and resonant searches. In (b), a resonance mass of 300~GeV is assumed. The gray hatched bands represent the uncertainties on the total backgrounds. These uncertainties are largely correlated from bin to bin.}
\label{fig:bbtt-SR}
\end{figure}

Systematic uncertainties from the trigger, luminosity, object identification, background estimate as well as Monte Carlo modeling of signal and background processes are taken into account in the background estimates and the calculation of signal yields.  
The impact of these systematic uncertainties varies for different background components and event categories. For the most sensitive $n_b\ge 2$ categories, the main background contributions are from top quark, fake $\tauhad$, and $Z\!\to\!\tau\tau$. The jet energy scale and resolution is the largest uncertainty for the top-quark contribution, ranging between 10\% and 19\% for the nonresonant and resonant searches. The leading source of systematic uncertainty for the fake $\tauhad$ background  is  the ``fake-factor'' determination, due to the uncertainties of the sample composition.
Varying the composition of $W$+jets, $Z$+jets, top quark and multijet events in the control samples by $\pm 50\%$ leads to a change in the estimated fake $\tauhad$ background by 9.5\%. The most important source of systematic uncertainty for the $Z\!\to\!\tau\tau$ background is the $t\bar{t}$ subtraction from the $Z\!\to\!\mu\mu$ sample used for the embedding, due to the uncertainty on the $t\bar{t}$ normalization. Its effect ranges from 8\% to 15\%. The overall systematic uncertainties on the total background contributions to the high (low) $\pTtt$ category of $n_b\ge 2$ are 12\% (9\%) for the nonresonant search and  14\% (14\%) for the resonant search. The largest contributions are from jet and tau energy scales and $b$-tagging. The modeling of top quark production is also an important systematic uncertainty for the category with two or more $b$-tagged jets and high $\pTtt$.

The uncertainties on the signal acceptances are estimated from experimental as well as theoretical sources. 
The total experimental systematic uncertainties vary between 12\% and 24\% for the categories with two or more $b$-tagged jets, and are dominated by the jet and tau energy scales and $b$-tagging. Theoretical uncertainties arise from the choice of parton distribution functions, the renormalization and factorization scales as well as the value of strong coupling constant $\alpha_{\rm s}$ used to generate the signal events. Uncertainties of 3\%, 1\% and 3\% from the three sources, respectively, are assigned to all signal acceptances.

For the nonresonant search, the observed ditau mass distribution agrees well with that of the estimated background events as shown in Fig.~\ref{fig:bbtt-SR}(a). For the resonant search, a small deficit with a local significance of approximately $2\sigma$ is observed in the data relative to the background expectation at $m_{bb\tau\tau}\sim 300$~GeV as is shown in Fig.~\ref{fig:bbtt-SR}(b). 
No evidence of Higgs boson pair production is present in the data. The resulting upper limits on Higgs boson pair production from these searches are described in Sec.~\ref{sec:results}.

\section{$\yyWW$}
\label{sec:yyWW}

This section describes the search for Higgs boson pair production in the $\yyWW$ decay channel, where one Higgs boson decays to a pair of photons and the other decays to a pair of $W$ bosons. The $h\!\to\!\gamma\gamma$ decay
is well suited for tagging the Higgs boson. The small Higgs boson width together with the excellent detector resolution for the diphoton mass strongly suppresses background contributions. Moreover, the $h\!\to\! WW^*$ decay has the largest branching ratio after $h\!\to\! \bb$. To reduce multijet backgrounds, one of the $W$ bosons is required to decay to an electron or a muon (either directly or through a tau lepton) whereas the other is required to decay hadronically, leading to the $\gamma\gamma \ell\nu qq'$ final state.

The data used in this analysis were recorded with diphoton triggers with an efficiency close to 100\% for diphoton events passing the final offline selection. The diphoton selection follows closely that of the ATLAS measurement of the $h\!\to\!\gamma\gamma$ production rate~\cite{atlas:2014hgg} and that of the $\bbyy$ analysis~\cite{Aad:2014yja}. Events are required to have two or more identified photons with the leading and subleading photon candidates having $\pT/\myy>0.35$ and 0.25, respectively, where $\myy$ is the invariant mass of the two selected photons. Only events with $\myy$ in the range of $105<\myy<160$~GeV are considered. 

Additional requirements are applied to identify the  $h\!\to\! WW^*\!\to\!\ell\nu qq'$ decay signature. Events must have two or more jets, and exactly one lepton, satisfying the identification criteria described in Sec.~\ref{sec:objects}. 
To reduce multijet backgrounds, the events are required to have $\met$ with significance greater than one.
Events with any $b$-tagged jet are vetoed to reduce contributions from top quark production. 

A total of 13 events pass the above selection. The final $\yyWW$ candidates are selected by requiring the diphoton mass $\myy$ to be within a $\pm 2\sigma$ window of the  Higgs boson mass in $h\!\to\!\gamma\gamma$ where $\sigma$ is taken to be 1.7~GeV. Due to the small number of events, both nonresonant and resonant searches proceed as counting experiments. 
The selection efficiency for the  $hh\!\to\! \gamma\gamma WW^*$ signal of SM nonresonant Higgs boson pair production is estimated using simulation to be 2.9\%. For the resonant production, the corresponding efficiency varies from 1.7\% at 260~GeV to 3.3\% at 500~GeV. These efficiencies include the branching ratios of the $W$ boson decays, but not those of the Higgs boson decays.

The background contributions considered are single SM Higgs boson production (gluon fusion, vector-boson fusion and associated production of $Wh$, $Zh$ and $t\bar{t}h$) and continuum backgrounds in the $\myy$ spectrum. Events from single Higgs boson production can mimic the $\yyWW$ signal if, for example, the Higgs boson decays to two photons and the rest of the event satisfies the $h\!\to\! WW^*\!\to\! \ell\nu qq'$ identification. These events would exhibit a diphoton mass peak at $m_h$. As in the $\bbtt$ analysis, their contributions are estimated from simulation using the SM cross sections~\cite{Heinemeyer:2013tqa}. 
The systematic uncertainty on the total yield of these backgrounds is estimated to be 29\%, dominated by the modeling of jet production~(27\%).
The total number of events expected from single SM Higgs production is therefore $0.25\pm 0.07$ with contributions of 0.14, 0.08 and 0.025 events from $Wh$, $t\bar{t}h$ and $Zh$ processes, respectively. Contributions from gluon and vector-boson fusion processes are negligible.

\begin{figure}[!h!tpb]
  \centering
\subfloat[]{\includegraphics[width=0.5\textwidth]{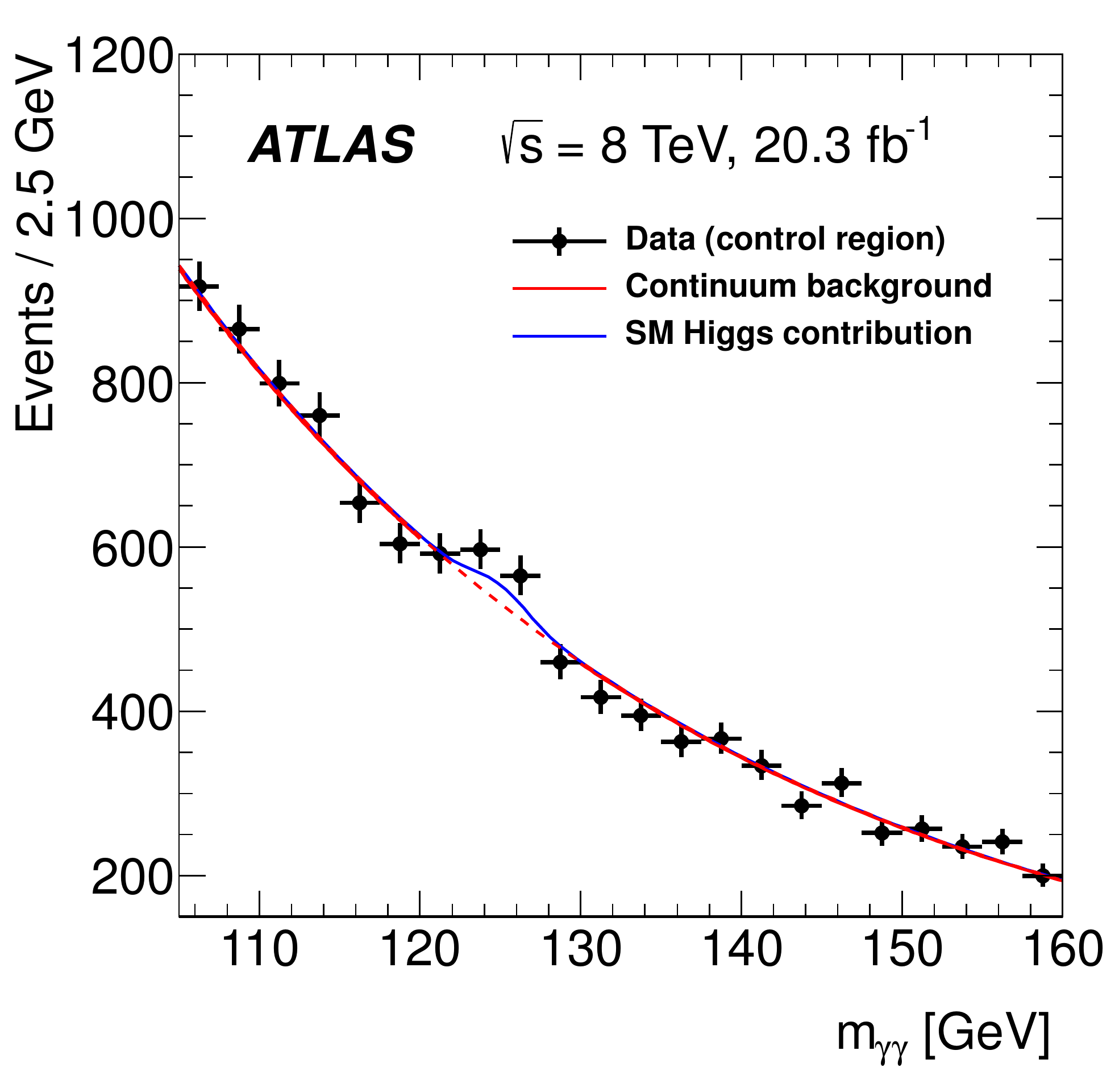}}
\subfloat[]{\includegraphics[width=0.5\textwidth]{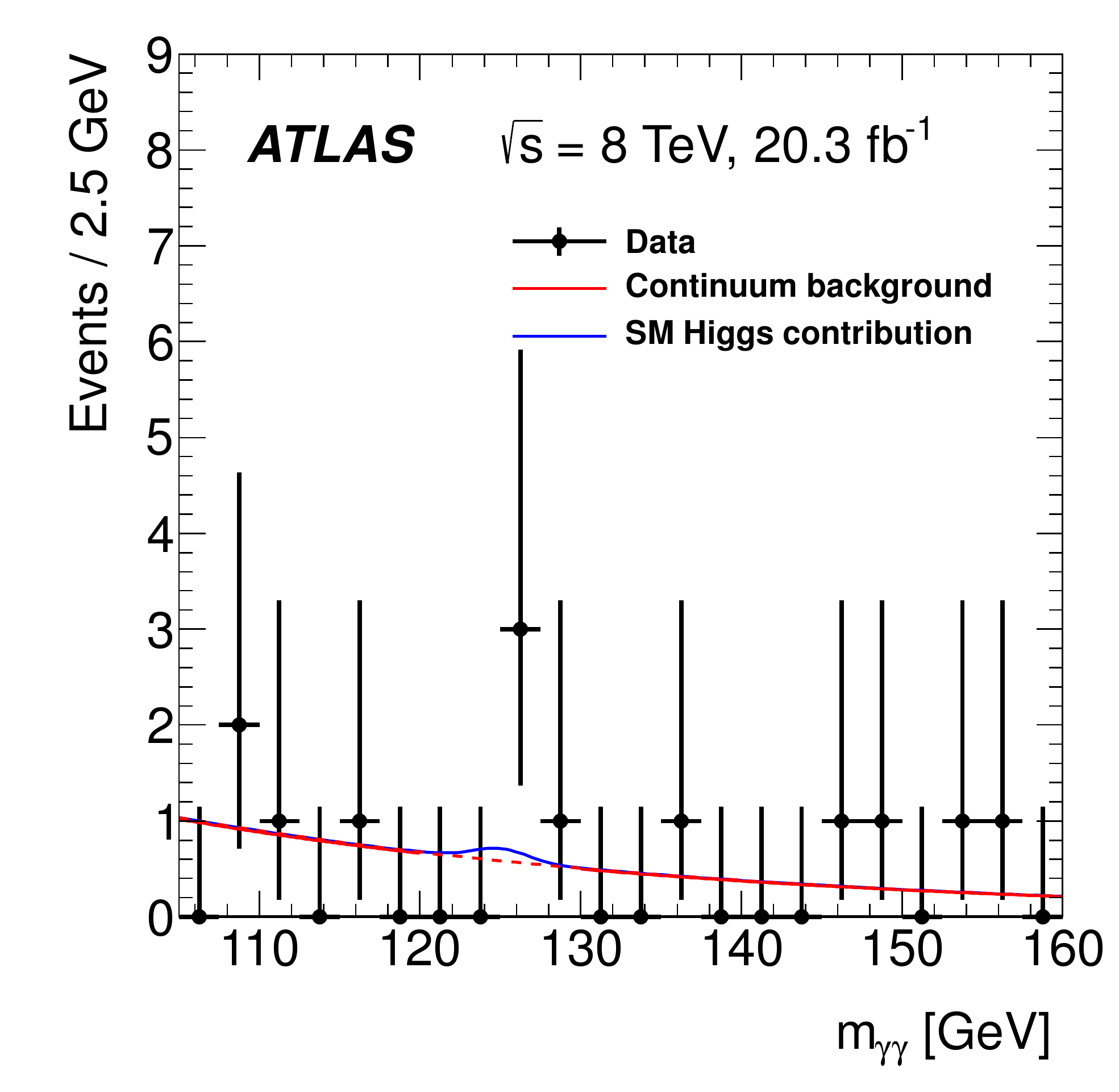}}
\caption{The distribution of the diphoton invariant mass for events passing (a) the relaxed requirements and (b) the final selection. The relaxed requirements include all final selections except those on the lepton and $\met$. The red curves represent the continuum background contributions and the blue curves include the contributions expected from single SM Higgs boson production estimated from simulation. The continuum background contributions in the signal $\myy$ mass window are shown as dashed lines.}
\label{fig:yyWW-myy}
\end{figure}

The background that is nonresonant in the $\gamma\gamma$ mass spectrum is measured using the continuum background in the $\myy$ spectrum. The major source of these backgrounds is $W\gamma\gamma+{\rm jets}$ events with a $W\!\to\! \ell\nu$ decay. 
 These events are expected to have a diphoton mass distribution with no resonant structure at $m_h$ and their contribution ($N_{\rm SR}^{\rm est}$) in the signal region, $\myy\in m_h\pm 2\sigma$, is estimated from the $\myy$ sidebands in the data:
$$N_{\rm SR}^{\rm est} = N_{\rm SB}^{\rm Data} \times \frac{f_{\rm SB}}{1-f_{\rm SB}}.$$
Here $N_{\rm SB}^{\rm Data}$ is the number of events in the data sidebands, defined as the mass region $105<\myy<160$~GeV excluding the signal region. The quantity $f_{\rm SB}$ is the fraction of background events in $105<\myy<160$~GeV falling into the signal mass window, and can in principle be determined from a fit of the observed $\myy$ distribution to an ansatz function. However, the small number of events after the final selection makes such a fit unsuitable. Instead, $f_{\rm SB}$ is determined in a data control sample, selected as the signal sample without the lepton and $\MET$ requirements.  Figure~\ref{fig:yyWW-myy}(a) shows the $\myy$ distribution of events in the control sample.  For the fit, an exponential function is used to model the sidebands and a wider region of $m_h\pm 5$~GeV is excluded to minimize potential signal contamination in the sidebands. The fit yields a value of $f_{\rm SB}=0.1348\pm 0.0001$. Varying the fit range of the sidebands leads to negligible changes. Different fit functions, such as a second-order polynomial or an exponentiated second-order polynomial, lead to a difference of 1.4\% in $f_{\rm SB}$.  To study the sample dependence of $f_{\rm SB}$, the fit is repeated for the control sample without the jet and $\met$ requirements and a difference of only 2\% is observed. Simulation studies show that the continuum background is dominated by $W(\ell\nu)\gamma\gamma+{\rm jets}$ production. The $\yylvjj$ events generated using MadGraph reproduce well the observed $\myy$ distribution. The potential difference between $\yyjj$ and $\yylvjj$ samples is studied using simulation. A difference below 1\% is observed. Taking all these differences as systematic uncertainties, the fraction of background events in the signal mass window is $f_{\rm SB} = 0.135\pm 0.004$.  With 9 ($N_{\rm SB}^{\rm Data}$) events observed in the data sidebands, it leads to $N^{\rm est}_{\rm SR}=1.40\pm 0.47$ events from the continuum background.  Figure~\ref{fig:yyWW-myy}(a) also shows the contribution expected from single SM Higgs boson production. The data prefer a larger cross section than the SM prediction for single Higgs boson production, consistent with the measurement reported in Ref.~\cite{Aad:2014lwa}.

The uncertainties on the signal acceptances are estimated following the same procedure as the $\bbtt$ analysis. The total experimental uncertainty is found to vary between 4\% and 7\% for different signal samples under consideration, dominated by the contribution from the jet energy scale. The theoretical uncertainties from PDFs, the renormalization and factorization scales, and the strong coupling constant are 3\%, 1\%, and 3\%, respectively, the same as for the $\bbtt$ analysis. 

The $\myy$ distribution of the selected events in the data is shown in Fig.~\ref{fig:yyWW-myy}(b). 
In total, 13 events are found with $105<\myy<160$~GeV. Among them, 4 events are in the signal mass window of $m_h\pm 2\sigma$  compared with $1.65\pm 0.47$ events expected from single SM Higgs boson production and continuum background processes.
The $p$-value of the background-only hypothesis is 3.8\%, corresponding to 1.8 standard deviations. 

 Assuming a cross section of 1~pb ($\sigma(gg\!\to\! hh)$ or $\sigma(gg\!\to\! H)\times {\rm BR}(H\!\to\! hh)$) for Higgs boson pair production, the expected number of signal events is $0.64\pm 0.05$ for the nonresonant production. For the resonant production, the corresponding numbers of events are $0.47\pm 0.05$ and $0.72\pm 0.06$ for a resonance mass of 300~GeV and 500~GeV, respectively. The implications of the search for Higgs boson pair production are discussed in Sec.~\ref{sec:results}.

\section{Combination procedure}
\label{sec:procedure}

The statistical analysis of the searches is based on the framework described in Refs.~\cite{Aad:2012an,Moneta:2010pm,Verkerke:2003ir,Cowan:2010js}. Profile-likelihood-ratio test statistics are used to measure the compatibility of the background-only hypothesis with the observed data and to test the hypothesis of Higgs boson pair production with its cross section as the parameter of interest.  Additional nuisance parameters are included to take into account systematic uncertainties and their correlations. The likelihood is the product of terms of the Poisson probability constructed from the final discriminant and of nuisance parameter constraints with either Gaussian, log-normal, or Poisson distributions. Upper limits on the Higgs boson pair production cross section are derived using the ${\rm CL}_s$ method~\cite{Read:2002hq}. 
For the combinations, systematic uncertainties that affect two or more analyses (such as those of luminosity, jet energy scale and resolutions, $b$-tagging, etc.) are modeled with common nuisance parameters.

For the $\bbtt$ analysis, Poisson probability terms are calculated for the four categories separately from the mass distributions of the ditau system for the nonresonant search (Fig.~\ref{fig:bbtt-SR}(a)) and of the $bb\tau\tau$ system for the resonant search (Fig.~\ref{fig:bbtt-SR}(b)). The $\mbbtt$ distributions of the resonant search are rebinned to ensure a sufficient number of events for the background prediction in each bin, in particular a single bin is used for $\mbbtt\gtrsim 400$~GeV for each category. For the $\yyWW$ analysis, event yields are used to calculate Poisson probabilities without exploiting shape information. 
The $\bbyy$ and $\bbbb$ analyses are published separately in Refs.~\cite{Aad:2014yja,Aad:2015uka}. However, the results are quoted at slightly different values of the Higgs boson mass $m_h$ and, therefore, have been updated using a common mass value of $m_h=125.4$~GeV~\cite{Aad:2014aba} for the combinations. The decay branching ratios of the Higgs boson $h$ and their uncertainties used in the combinations are taken from Ref.~\cite{Heinemeyer:2013tqa}. Table~\ref{tab:overview} is a summary of the number of categories and final discriminants used for each analysis.

\begin{table}[htb]
\caption{An overview of the number of categories and final discriminant distributions used for both the nonresonant and resonant searches. Shown in the last column are the mass ranges of the resonant searches. }
\begin{center}
\begin{tabular}{cccccccc}\hline\hline
& $hh$         & \multicolumn{2}{c}{Nonresonant search} && \multicolumn{3}{c}{Resonant search}  \\ \cline{3-4}\cline{6-8} 
& final state  & Categories    &   Discriminant    &&   Categories   &   Discriminant    & $m_H$ [GeV] \\ \hline
& $\gamma\gamma b\bar{b}$   &   1    &  $\myy$           &&      1    & event yields     &  260--500 \\
& $\gamma\gamma WW^*$       &   1    &  event yields     &&      1    & event yields     &  260--500 \\
& $b\bar{b}\tau\tau$        &   4    &  $m_{\tau\tau}$   &&      4    & $m_{bb\tau\tau}$ &  260--1000 \\
& $b\bar{b}b\bar{b}$        &   1    &  event yields     &&      1    & $m_{bbbb}$       &  500--1500 \\
\hline\hline
\end{tabular}
\end{center}
\label{tab:overview}
\end{table}

The four individual analyses are sensitive to different kinematic regions of the $hh$ production and decays. The combination is performed assuming that the relative contributions of these regions to the total cross section are modeled by the {\sc MadGraph5}~\cite{Alwall:2011uj} program used to simulate the $hh$ production.

\section{Results}
\label{sec:results}

In this section, the limits on the nonresonant and resonant searches are derived.  The results of the $\bbtt$ and $\yyWW$ analyses are first determined and then combined with previously published results of the $\bbyy$ and $\bbbb$ analyses. The impact of the leading systematic uncertainties is also discussed. 

The observed and expected upper limits at 95\% CL on the cross section of nonresonant production of a Higgs boson pair are shown in Table~\ref{tab:NRSlimits}. These limits are to be  compared with the SM prediction of $9.9\pm 1.3$~fb~\cite{deFlorian:2013jea} for $gg\!\to\! hh$ production with $m_h=125.4$~GeV. Only the gluon fusion production process is considered.
The observed (expected) cross-section limits are 1.6 (1.3)~pb and 11.4 (6.7)~pb from the $\bbtt$ and $\yyWW$ analyses, respectively. Also shown in the table are the cross-section limits relative to the SM expectation. The results are combined with those of the $\bbyy$ and $\bbbb$ analyses. The $p$-value of compatibility of the combination with the SM hypothesis is 4.4\%, equivalent to 1.7 standard deviations. The low $p$-value is a result of the  excess of events observed in the $\bbyy$ analysis. The combined observed (expected) upper limit on $\sigma(gg\!\to\! hh)$ is 0.69~(0.47)~pb, corresponding to 70~(48) times the cross section predicted by the SM. The $\bbbb$ analysis has the best expected sensitivity followed by the $\bbyy$ analysis. The observed combined limit is slightly weaker than that of the $\bbbb$ analysis, largely due to the aforementioned excess.

\begin{table}[htbp]
\caption{The expected and observed 95\% CL upper limits on the cross sections of nonresonant 
         $gg\!\to\! hh$ production at $\sqrt{s}=8$~TeV from individual analyses and their combinations. 
         SM values are assumed for the $h$ decay branching ratios. The cross-section limits normalized 
         to the SM value are also included. }\vspace*{-0.2cm}
\begin{center}
\begin{tabular}{lccccc}\hline\hline
  Analysis   & \hsc $\gamma\gamma  \bb$ \hsc &  $\gamma\gamma WW^*$  & \hsc $\bb\tau\tau$ \hsc &  $\bb\bb$  & \hsa Combined \hsa \tsp \\ \hline \\[-3mm]
             & \multicolumn{5}{c}{Upper limit on the cross section [pb]} \\ \hline
  Expected   &  1.0  &  6.7  &  1.3  & 0.62  &  0.47 \\
  Observed   &  2.2  &  11   &  1.6  & 0.62  &  0.69 \\  \hline \\[-3mm]
             & \multicolumn{5}{c}{Upper limit on the cross section relative to the SM prediction} \\ \hline
  Expected   & 100 &  680 & 130 &   63 & 48 \\
  Observed   & 220 & 1150 & 160 &   63 & 70 \\
\hline\hline
\end{tabular}
\end{center}
\label{tab:NRSlimits}
\end{table}

The impact of systematic uncertainties on the cross-section limits is studied using the signal-strength parameter $\mu$, defined as the ratio of the extracted to the assumed signal cross section (times branching ratio ${\rm BR}(H\!\to\! hh)$ for the resonant search). The resulting shifts in $\mu$ depend on the actual signal-strength value. For illustration, they are evaluated using a cross section of 1~pb for $gg\!\to\!(H\!\to\!) hh$, comparable to the limits set.
The effects of the most important uncertainty sources are shown in Table~\ref{tab:systematics}.  The leading contributions are from the background modeling, $b$-tagging, the $h$ decay branching ratios, jet and $\met$ measurements. 
The large impact of the $b$-tagging systematic uncertainty reflects the relatively large weight of the $\bbbb$ analysis in the combination. 
For the $\bbtt$ analysis alone, the three leading systematic sources are the background estimates, jet and $\met$ measurements, and lepton and $\tauhad$ identifications. For the $\yyWW$ analysis, they are the background estimates, jet and $\met$ measurements and theoretical uncertainties of the decay branching ratios of the Higgs boson $h$. 

\begin{table}
\caption{The impact of the leading systematic uncertainties on the signal-strength parameter $\mu$ of a hypothesized signal for both the nonresonant and resonant ($m_H=300,\, 600$~GeV) searches. For the signal hypothesis, a Higgs boson pair production cross section ($\sigma(gg\!\to\! hh)$ or $\sigma(gg\!\to\! H)\times {\rm BR}(H\!\to\! hh)$) of 1~pb is assumed. }
\begin{center}
\scalebox{0.95}{
\begin{tabular}{lcclcclc}\hline\hline
\multicolumn{2}{c}{Nonresonant search} && \multicolumn{5}{c}{Resonant search} \\ \cline{1-2}\cline{4-8}
       &                      & & \multicolumn{2}{c}{$m_H=300$ GeV} && \multicolumn{2}{c}{$m_H=600$ GeV}  \\ \cline{4-5}\cline{7-8} 
Source & $\Delta\mu/\mu$ [\%] &      & Source & $\Delta\mu/\mu$ [\%]     && Source & $\Delta\mu/\mu$ [\%] \\ \hline 
Background model &   11  && Background model      &  15    && $b$-tagging      &   10  \\
$b$-tagging      &  7.9  && Jet and $\met$        & 9.9    && $h$ BR           &  6.3  \\
$h$ BR           &  5.8  && Lepton and $\tauhad$  & 6.9    && Jet and $\met$   &  5.5  \\
Jet and $\met$   &  5.5  && $h$ BR                & 5.9    && Luminosity       &  2.7  \\
Luminosity       &  3.0  && Luminosity            & 4.0    && Background model &  2.4  \\ \hline 
Total            &   16  && Total                 & 21     && Total            &   14  \\
\hline\hline
\end{tabular}}
\end{center}
\label{tab:systematics}
\end{table}

For the resonant production, limits are set on the cross section of $gg\!\to\! H$ production of the heavy Higgs boson times its branching ratio ${\rm BR}(H\!\to\! hh)$ as a function of the heavy Higgs boson mass $m_H$. The observed (expected) limits of the $\bbtt$ and $\yyWW$ analyses are illustrated in Fig.~\ref{fig:limits} and listed in Table~\ref{tab:resonant} (along with results from the $\bbyy$ and $\bbbb$ analyses). The $m_H$ search ranges  are 260--1000~GeV for $\bbtt$ and 260--500~GeV for $\yyWW$. For the $\bbtt$ analysis, the observed limit around $m_H\sim 300$~GeV is considerably lower than the expectation, reflecting the deficit in the observed $m_{bb\tau\tau}$ distribution. At high mass, the limits are correlated since a single bin is used for $m_{bb\tau\tau}\gtrsim 400$~GeV. The decrease in the limit as $m_H$ increases is a direct consequence of increasing selection efficiency for the signal. This is also true for the $\yyWW$ analysis as the event selection is independent of $m_H$.

\begin{figure}[!h!tpb]
  \centering
\subfloat[]{\includegraphics[width=0.5\textwidth]{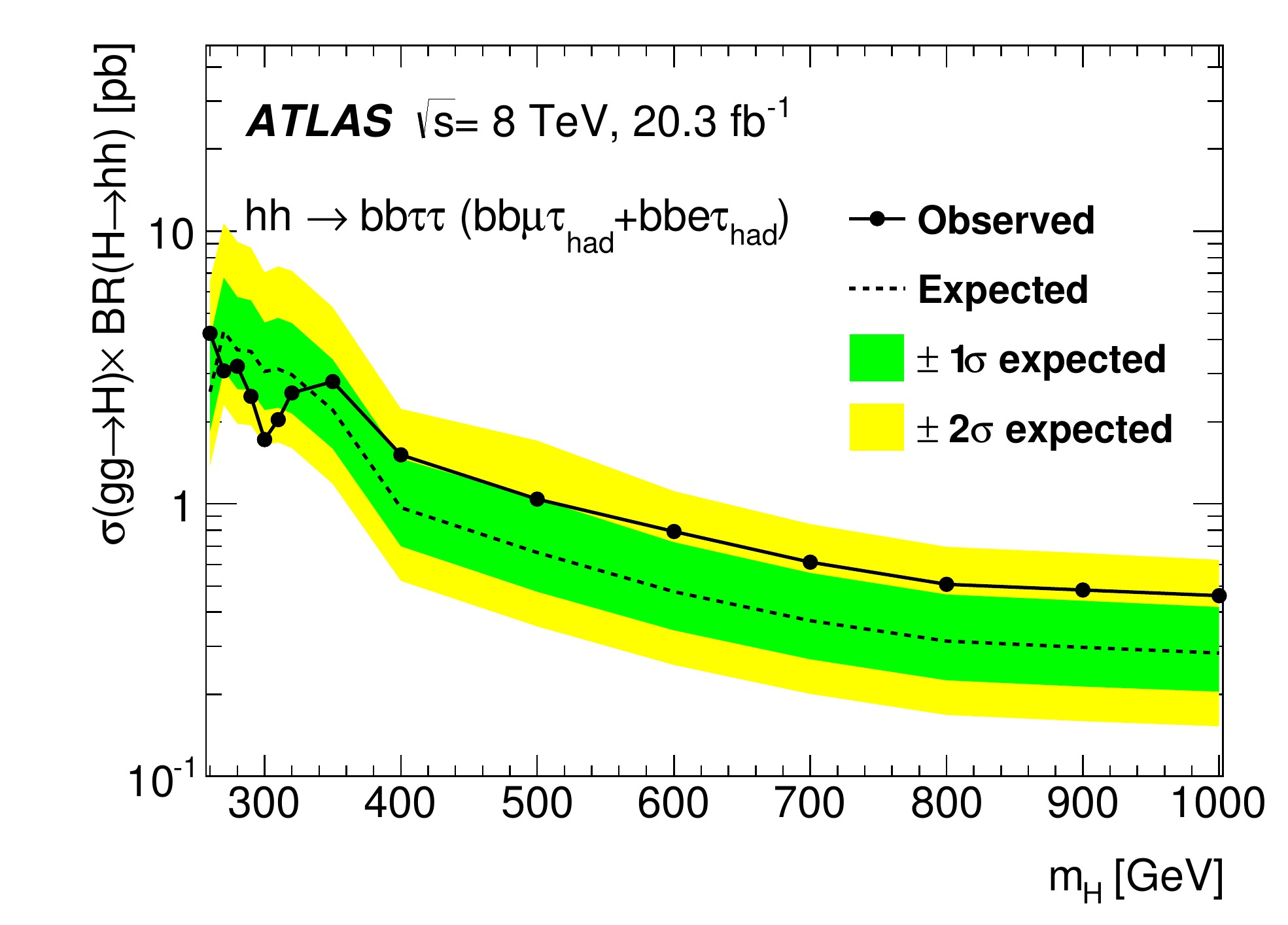}\label{fig:limits-bbtt}}
\subfloat[]{\includegraphics[width=0.5\textwidth]{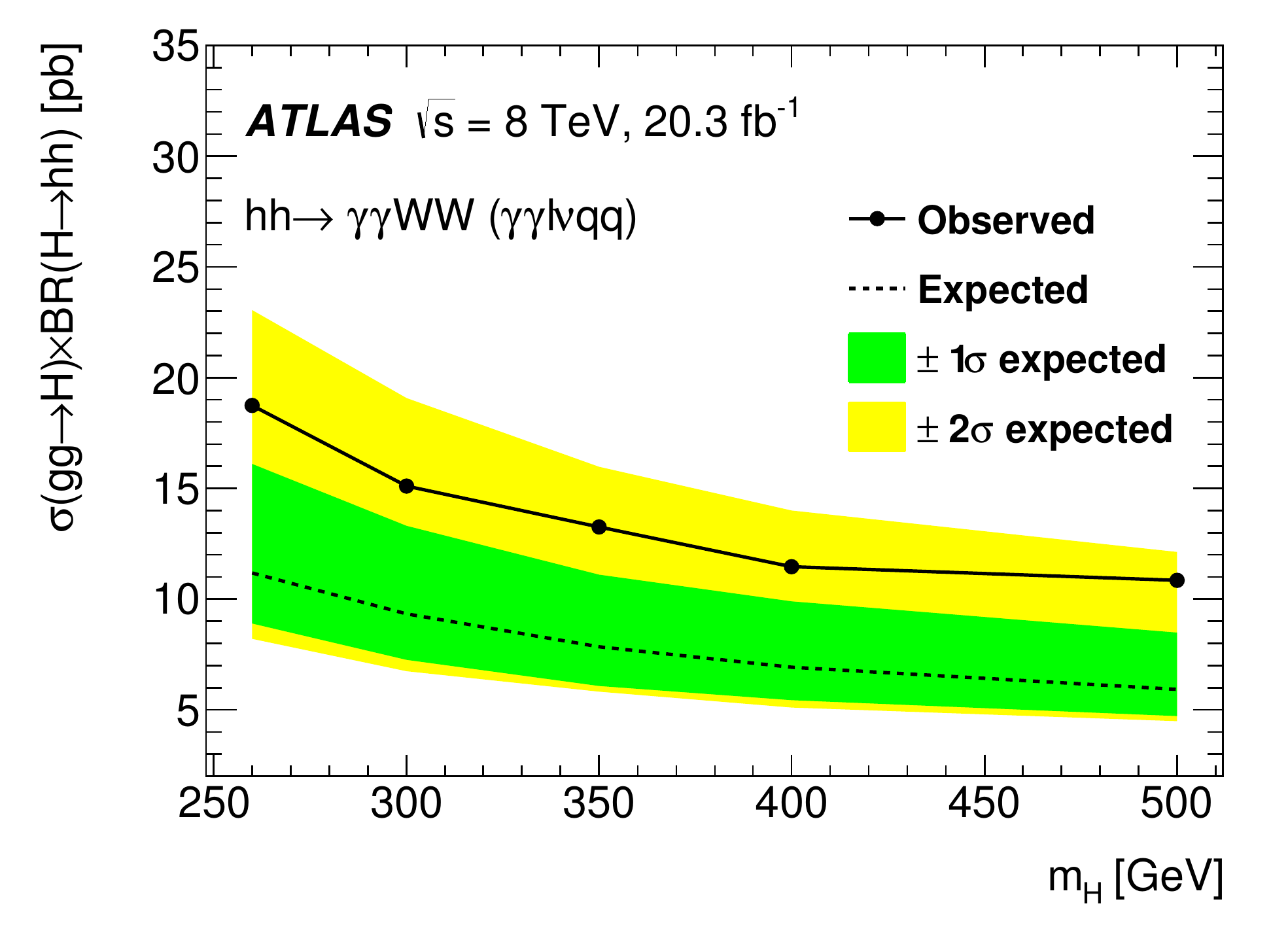}\label{fig:limits-yyWW}}
  \caption{The observed and expected upper limit at 95\% CL on $\sigma(gg\!\to\! H)\times{\rm BR}(H\!\to\! hh)$ at $\sqrt{s}=8$~TeV as a function of $m_H$ from the resonant (a) $\bbtt$ and (b) $\yyWW$ analyses. The search ranges of the resonance mass are 260--1000~GeV for $\bbtt$ and 260--500~GeV for $\yyWW$. The green and yellow bands represent $\pm 1\sigma$ and $\pm 2\sigma$ ranges on the expected limits, respectively. }
\label{fig:limits}
\end{figure}

\begin{table}[htb]
\caption{The expected and observed 95\% CL upper limits on  $\sigma(gg\!\to\! H)\times {\rm BR}(H\!\to\! hh)$ 
         in pb at $\sqrt{s}=8$~TeV from individual analyses and their combinations. 
         The SM branching ratios are assumed for the light Higgs boson decay. }\vspace*{-0.2cm}
\begin{center}{\small
\begin{tabular}{cccccccccccc}\hline\hline
$m_H$      & \multicolumn{5}{c}{Expected limit [pb]}   && \multicolumn{5}{c}{Observed limit [pb]} \\ \cline{2-6} \cline{8-12}
[GeV]        & $\gamma\gamma \bb$ & $\gamma\gamma WW^*$ & $\bb \tau\tau$ & $\bb\bb$ & Combined &
             & $\gamma\gamma \bb$ & $\gamma\gamma WW^*$ & $\bb \tau\tau$ & $\bb\bb$ & Combined \\ \hline
 260  & 1.70 & 11.2 &  2.6 & -- &  1.1 && 2.29 & 18.7 & 4.2 & -- & 2.1 \\
 300  & 1.53 &  9.3 &  3.1 & -- &  1.2 && 3.54 & 15.1 & 1.7 & -- & 2.0 \\
 350  & 1.23 &  7.8 &  2.2 & -- & 0.89 && 1.44 & 13.3 & 2.8 & -- & 1.5 \\
 400  & 1.00 &  6.9 & 0.97 & -- & 0.56 && 1.00 & 11.5 & 1.5 & -- & 0.83 \\
 500  & 0.72 &  5.9 & 0.66 & -- & 0.38 && 0.71 & 10.9 & 1.0 & -- & 0.61 \\ \\[-2mm]

 500  & -- & -- & 0.66 &  0.17 &  0.16 && -- & -- &  1.0 &  0.16 &  0.18 \\
 600  & -- & -- & 0.48 & 0.070 & 0.067 && -- & -- & 0.79 & 0.072 & 0.079 \\
 700  & -- & -- & 0.31 & 0.041 & 0.040 && -- & -- & 0.61 & 0.038 & 0.040 \\
 800  & -- & -- & 0.31 & 0.028 & 0.028 && -- & -- & 0.51 & 0.046 & 0.049 \\
 900  & -- & -- & 0.30 & 0.022 & 0.022 && -- & -- & 0.48 & 0.015 & 0.015 \\
1000  & -- & -- & 0.28 & 0.018 & 0.018 && -- & -- & 0.46 & 0.011 & 0.011 \\
\hline\hline
\end{tabular}}
\end{center}
\label{tab:resonant}
\end{table}

The $\bbyy$ and $\bbbb$ analyses are published separately and the mass range covered by the two analyses are 260--500~GeV and 500--1500~GeV, respectively. The results of these four analyses, summarized in Table~\ref{tab:resonant}, are combined for the mass range 260--1000~GeV assuming the SM values of the $h$ decay branching ratios.  To reflect the better mass resolutions of the $\bbbb$ and $\bbyy$ analyses, the combination is performed with smaller mass steps than those of the $\bbtt$ and $\yyWW$ analyses. 
The most significant excess in the combined results is at a resonance mass of 300~GeV with a local significance of $2.5\sigma$, largely due to the $3.0\sigma$ excess observed in the $\bbyy$ analysis~\cite{Aad:2014yja}. The upper limit on $\sigma(gg\!\to\! H)\times{\rm BR}(H\!\to\! hh)$ varies from 2.1~pb at 260 GeV to 0.011~pb at 1000~GeV. These limits are shown in Fig.~\ref{fig:combined} as a function of $m_H$. For the low-mass region of 260--500~GeV, 
both the $\bbyy$ and $\bbtt$ analyses contribute significantly to the combined sensitivities.
Above 500~GeV, the sensitivity is dominated by the $\bbbb$ analysis. Table~\ref{tab:systematics} shows the impact of the leading systematic uncertainties for a heavy Higgs boson mass of 300~GeV and 600~GeV. As in the nonresonant search, the systematic uncertainties with the largest impact on the sensitivity are from the uncertainties on the background modeling, $b$-tagging, jet and $\met$ measurements, and the $h$ decay branching ratios. 
These limits are directly applicable to models such as those of Refs.~\cite{Hill:1987ea,Binoth:1996au,Schabinger:2005ei,Patt:2006fw,Pruna:2013bma,Robens:2015gla} in which the Higgs boson $h$ has the same branching ratios as the SM Higgs boson.

\begin{figure}[!h!tpb]
  \centering
\includegraphics[width=0.85\textwidth]{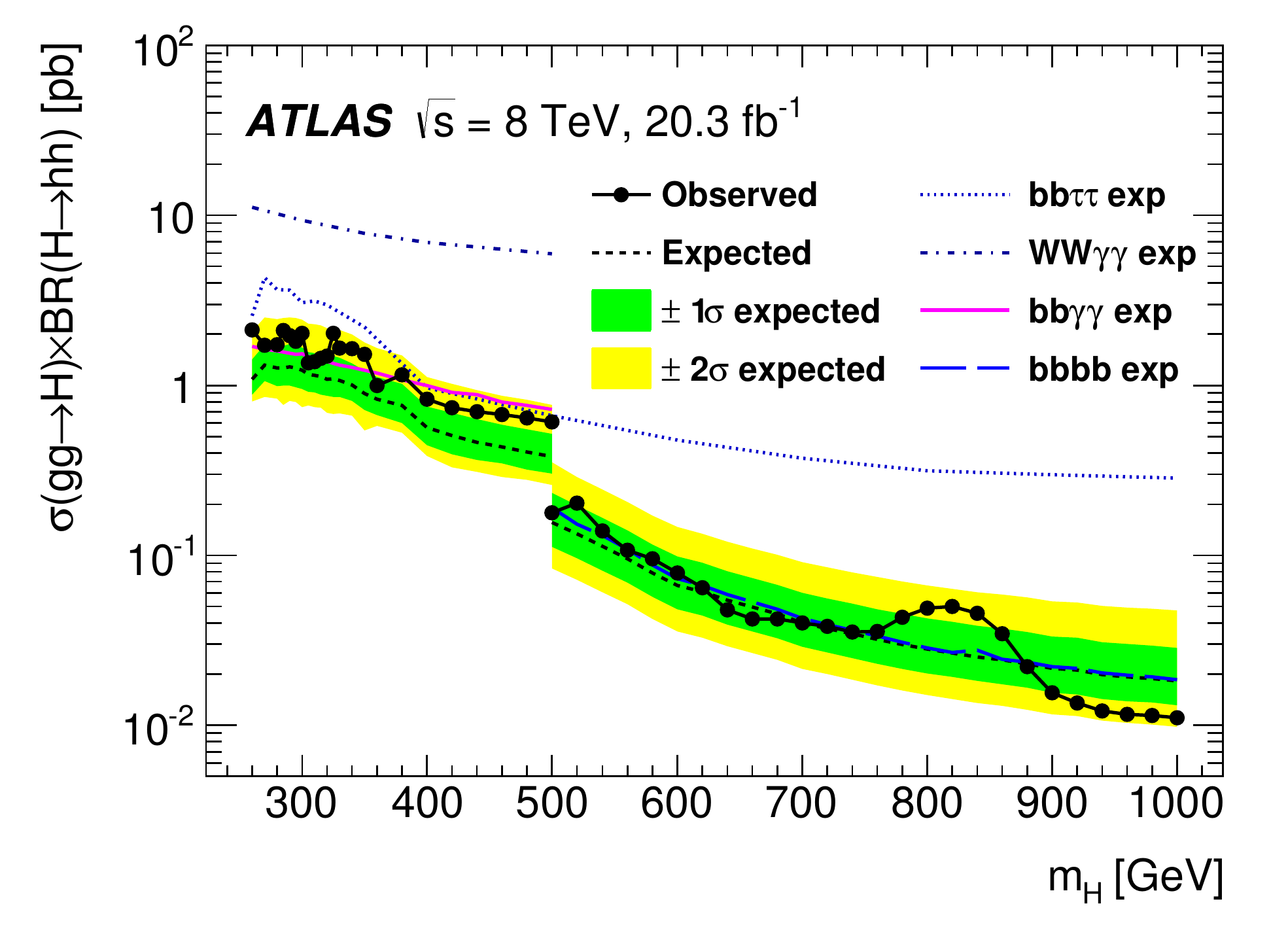}
  \caption{The observed and expected 95\% CL upper limits on $\sigma(gg\!\to\! H)\times{\rm BR}(H\!\to\! hh)$ 
           at $\sqrt{s}=8$~TeV as functions of the heavy Higgs boson mass $m_H$, combining resonant searches in 
           $hh\!\to\! \gamma\gamma bb,\ bbbb,\ bb\tau\tau$ and $\gamma\gamma WW^*$ final states.
           The expected limits from individual analyses are also shown. The combination assumes SM values
           for the decay branching ratios of the lighter Higgs boson $h$. The green and yellow 
           bands represent $\pm1\sigma$ and $\pm2\sigma$ uncertainty ranges of the expected combined limits. 
           The improvement above $m_H=500$~GeV is due to the sensitivity of the $\bbbb$ analysis. 
           The more finely spaced  mass points of the combination reflect the better mass resolutions
           of the $\bbyy$ and $\bbbb$ analyses than those of the $\bbtt$ and $\yyWW$ analyses.
  }
\label{fig:combined}
\end{figure}

\section{Interpretation}
\label{sec:interpretation}

The upper cross-section limits of the resonant search are interpreted in two MSSM scenarios, one referred to as the hMSSM~\cite{Djouadi:2013uqa,Djouadi:2015jea} and the other as the low-tb-high~\cite{LHCtbNote}. In the interpretation, the CP-even light and heavy Higgs bosons of the MSSM are assumed to be the Higgs bosons $h$ and $H$ of the search, respectively.  
The natural width of the heavy Higgs boson $H$ where limits are set in these scenarios is sufficiently smaller than the experimental resolution, which is at best 1.5\%, that its effect can be neglected.

In the hMSSM scenario, the mass of the light CP-even Higgs boson is fixed to 125 GeV in the whole parameter space. This is achieved by implicitly allowing the supersymmetry-breaking scale $m_{\rm S}$ to be very large, which is especially true in the low $\tan\beta$ region where $m_{\rm S}\! \gg\! 1$~TeV, and making assumptions about the CP-even Higgs boson mass matrix and its radiative corrections, as well as the Higgs boson coupling dependence on the MSSM parameters. Here $\tan\beta$ is the ratio of the vacuum expectation values of the two doublet Higgs fields. 
The ``low-tb-high'' MSSM scenario follows a similar approach, differing in that explicit choices are made  for the supersymmetry-breaking parameters~\cite{LHCtbNote}. The mass of the light Higgs boson is not fixed in this scenario, but is approximately 125~GeV in most of the parameter space. The $m_h$ value grows gradually from 122~GeV at $m_A\sim 220$~GeV to 125~GeV as $m_A$ approaching infinity. 
Higgs boson production cross sections through the gluon-fusion process  are calculated with {\sc SusHi}~1.4.1~\cite{Harlander:2012pb,Harlander:2002wh,Aglietti:2004nj} for both scenarios. Higgs boson decay branching ratios are calculated with HDECAY~6.42~\cite{Djouadi:1997yw} following the prescription of Ref.~\cite{Djouadi:2015jea} for the hMSSM scenario and with FeynHiggs 2.10.0~\cite{Heinemeyer:1998yj,Degrassi:2002fi,Hahn:2013ria} for the low-tb-high scenario.

The upper limits on $\sigma(gg\to H)\times {\rm BR}(H\to hh)$ can be interpreted as exclusion regions in the $(\tan\beta, m_A)$ plane. In both scenarios, the Higgs boson pair production rate $\sigma(gg\to H)\times {\rm BR}(H\to hh)$ depends on $\tan\beta$ and the mass of the CP-odd Higgs boson ($m_A$), and so does the mass of the heavy CP-even Higgs boson $H$. The values of $m_A$ and $m_H$ are generally different: $m_H$ can be as much as 70~GeV above $m_A$ in the parameter space relevant for this publication with the difference in masses decreasing for increasing values of $\tan\beta$ or $m_A$. Constant $m_H$ lines for a few selected values are shown in Fig.~\ref{fig:interpr}. The decay branching ratios of the light Higgs boson in these scenarios depend on $\tan\beta$ and $m_A$ and are different from the corresponding SM values used to derive the upper limits shown in Table~\ref{tab:resonant}. 
The upper limits, as functions of $m_H$, are recomputed; the $hh$ decay fractions for each final state are fixed to their smallest value found in $1<\tan\beta<2$, the range of the expected sensitivity. This approach yields conservative limits, but simplifies the computation as the limit calculation does not have to be repeated at each $\tan\beta$ value. The results are used to set exclusions in the $(\tan\beta,\, m_A)$ plane as shown in Fig.~\ref{fig:interpr}. The analysis is sensitive to the region of low $\tan\beta$ and $m_A$ values in the range $\sim$ 200--350~GeV. For $m_A\lesssim 200$~GeV, $m_H$ is typically below the $2m_h$ threshold of the $H\to hh$ decay, whereas above 350~GeV, the $H\to hh$ decay is suppressed because of the dominance of the $H\to t\bar{t}$ decay. The observed exclusion region in the $(\tan\beta,m_A)$ plane is smaller than the expectation, reflecting the small excess observed in the data.

\begin{figure}[!h!tpb]
  \centering
\subfloat[]{\includegraphics[width=0.5\textwidth]{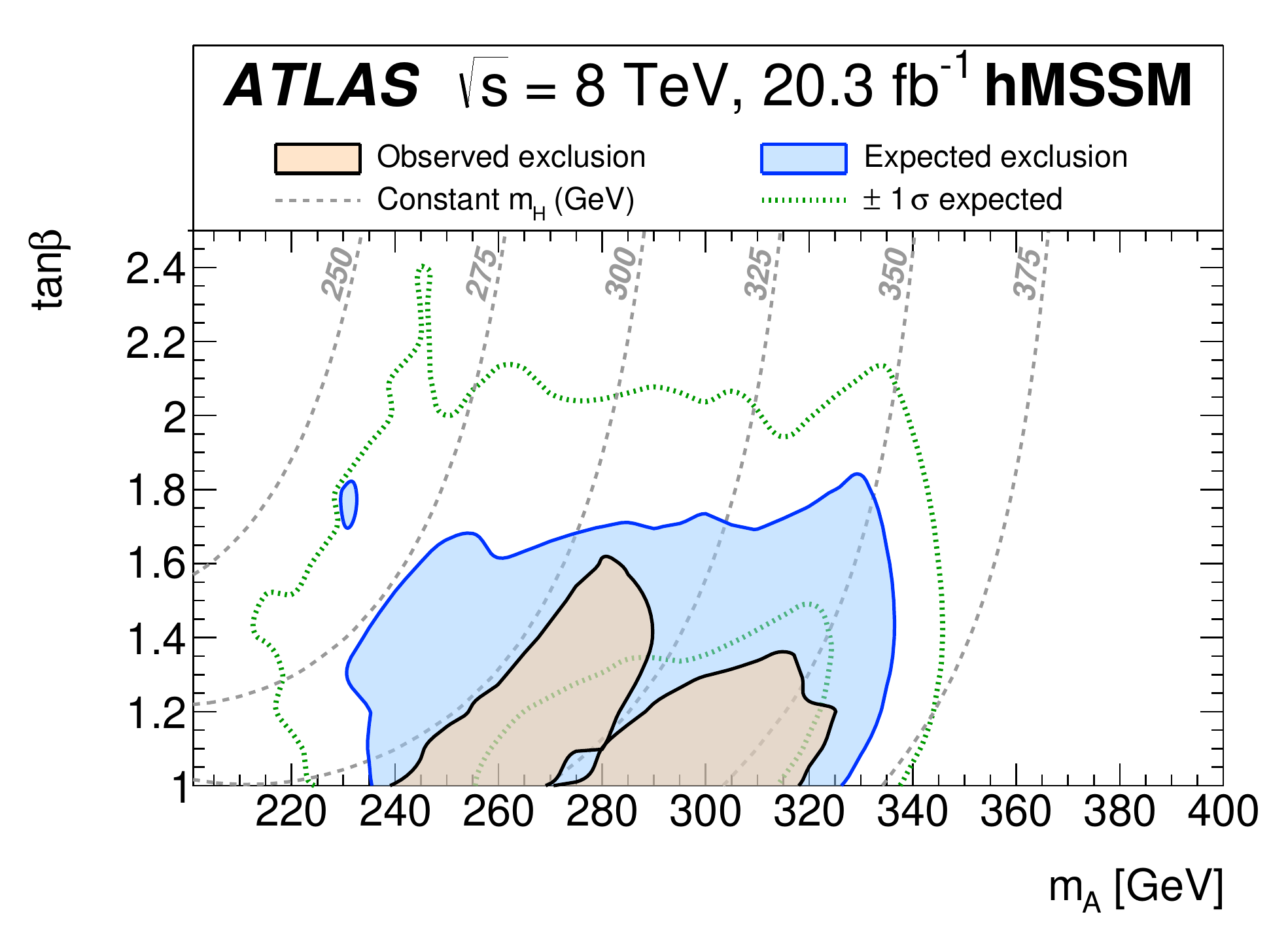}}
\subfloat[]{\includegraphics[width=0.5\textwidth]{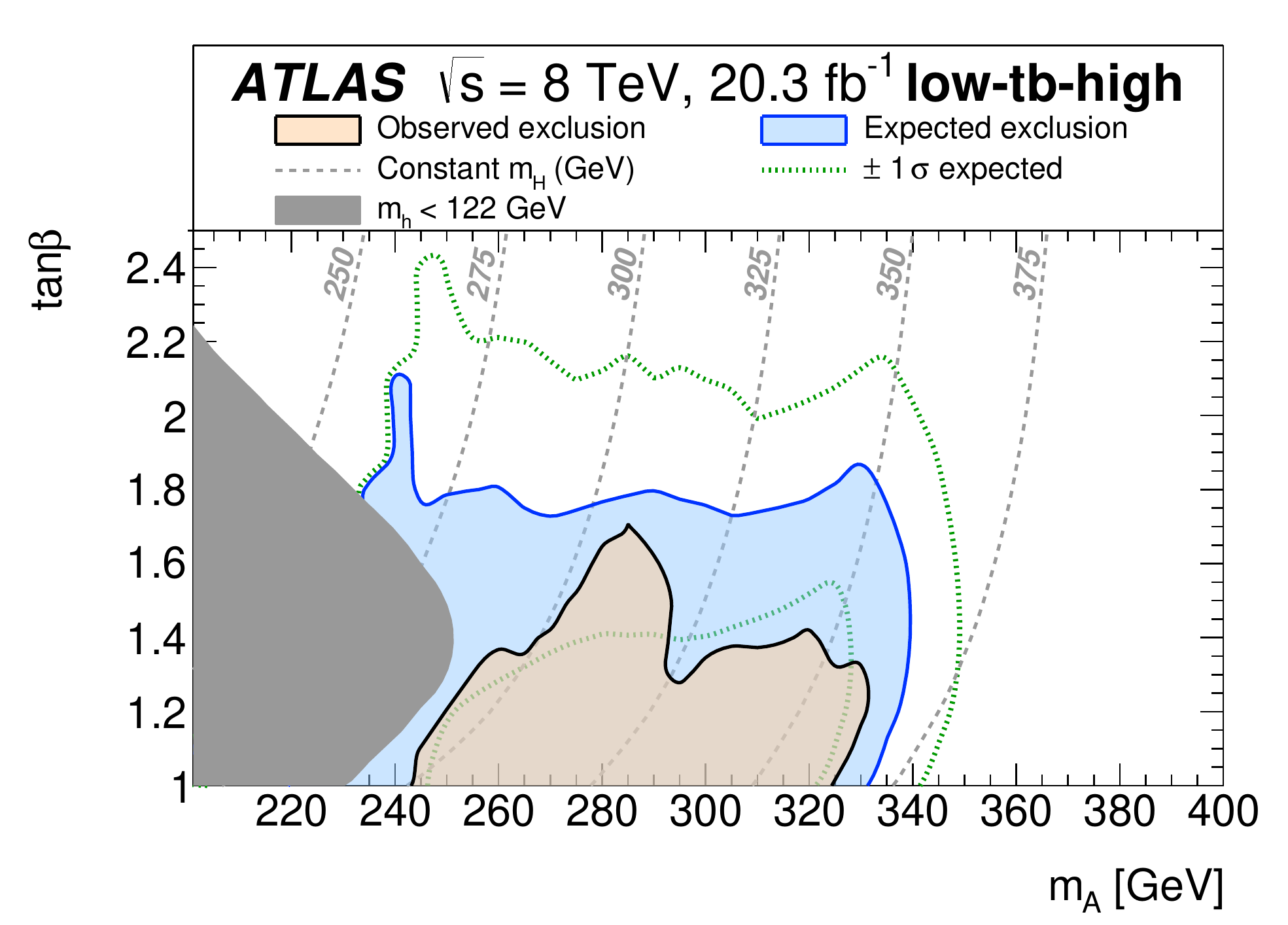}}
 \caption{The observed and expected 95\% CL exclusion regions in the $(\tan\beta, m_A)$ plane of MSSM scenarios from the resonant search: (a) the hMSSM scenario and (b) the low-tb-high scenario. The green dotted lines delimit the $\pm 1\sigma$ uncertainty ranges of the expected exclusion regions. The gray dashed lines show the constant values of the heavy CP-even Higgs boson mass. The improved sensitivity in the expected exclusion on the contour line of $m_H\sim 260$~GeV reflects the improved expected limit on the cross section while the hole or the wedge around the $m_H\sim 325$~GeV contour line in the observed exclusion is the result of a small excess at this mass, see Fig.~\ref{fig:combined}. The gray shaded region in (b) shows the region where the mass of the light CP-even Higgs boson is inconsistent with the measured value of 125.4~GeV. There is no such region in (a) by construction.}
\label{fig:interpr}
\end{figure}

\section{Summary}
\label{sec:summary}

This paper summarizes the search for both nonresonant and resonant Higgs boson pair production in proton--proton collisions from approximately 20~\ifb\ of data at a center-of-mass energy of 8 TeV recorded by the ATLAS detector at the LHC. The search is performed in $hh\!\to\! \bb\tau\tau$ and $\gamma\gamma WW^*$ final states. No significant excess is observed in the data beyond the background expectation. Upper limits on the $hh$ production cross section are derived. Combining with the $hh\!\to\! \gamma\gamma\bb,\, \bb\bb$ searches, a 95\% CL upper limit of 0.69~pb on the cross section of the nonresonant $hh$ production is observed compared with the expected limit of 0.47~pb. This observed upper limit is approximately 70 times the SM $gg\!\to\! hh$ production cross section. For the production of a narrow heavy resonance decaying to a pair of light Higgs bosons, the observed (expected) upper limit on $\sigma(gg\!\to\! H)\times {\rm BR} (H\!\to\! hh)$ varies from 2.1~(1.1)~pb at 260~GeV to 0.011~(0.018)~pb at 1000 GeV. These limits are obtained assuming SM values for the $h$ decay branching ratios. Exclusion regions in the parameter space of simplified MSSM scenarios are also derived.

\section*{Acknowledgments}

We thank CERN for the very successful operation of the LHC, as well as the
support staff from our institutions without whom ATLAS could not be
operated efficiently.

We acknowledge the support of ANPCyT, Argentina; YerPhI, Armenia; ARC, Australia; BMWFW and FWF, Austria; ANAS, Azerbaijan; SSTC, Belarus; CNPq and FAPESP, Brazil; NSERC, NRC and CFI, Canada; CERN; CONICYT, Chile; CAS, MOST and NSFC, China; COLCIENCIAS, Colombia; MSMT CR, MPO CR and VSC CR, Czech Republic; DNRF, DNSRC and Lundbeck Foundation, Denmark; IN2P3-CNRS, CEA-DSM/IRFU, France; GNSF, Georgia; BMBF, HGF, and MPG, Germany; GSRT, Greece; RGC, Hong Kong SAR, China; ISF, I-CORE and Benoziyo Center, Israel; INFN, Italy; MEXT and JSPS, Japan; CNRST, Morocco; FOM and NWO, Netherlands; RCN, Norway; MNiSW and NCN, Poland; FCT, Portugal; MNE/IFA, Romania; MES of Russia and NRC KI, Russian Federation; JINR; MESTD, Serbia; MSSR, Slovakia; ARRS and MIZ\v{S}, Slovenia; DST/NRF, South Africa; MINECO, Spain; SRC and Wallenberg Foundation, Sweden; SERI, SNSF and Cantons of Bern and Geneva, Switzerland; MOST, Taiwan; TAEK, Turkey; STFC, United Kingdom; DOE and NSF, United States of America. In addition, individual groups and members have received support from BCKDF, the Canada Council, CANARIE, CRC, Compute Canada, FQRNT, and the Ontario Innovation Trust, Canada; EPLANET, ERC, FP7, Horizon 2020 and Marie Skłodowska-Curie Actions, European Union; Investissements d'Avenir Labex and Idex, ANR, Region Auvergne and Fondation Partager le Savoir, France; DFG and AvH Foundation, Germany; Herakleitos, Thales and Aristeia programmes co-financed by EU-ESF and the Greek NSRF; BSF, GIF and Minerva, Israel; BRF, Norway; the Royal Society and Leverhulme Trust, United Kingdom.

The crucial computing support from all WLCG partners is acknowledged
gratefully, in particular from CERN and the ATLAS Tier-1 facilities at
TRIUMF (Canada), NDGF (Denmark, Norway, Sweden), CC-IN2P3 (France),
KIT/GridKA (Germany), INFN-CNAF (Italy), NL-T1 (Netherlands), PIC (Spain),
ASGC (Taiwan), RAL (UK) and BNL (USA) and in the Tier-2 facilities
worldwide.

\printbibliography

\clearpage

\begin{flushleft}
{\Large The ATLAS Collaboration}

\bigskip

G.~Aad$^{\rm 85}$,
B.~Abbott$^{\rm 113}$,
J.~Abdallah$^{\rm 151}$,
O.~Abdinov$^{\rm 11}$,
R.~Aben$^{\rm 107}$,
M.~Abolins$^{\rm 90}$,
O.S.~AbouZeid$^{\rm 158}$,
H.~Abramowicz$^{\rm 153}$,
H.~Abreu$^{\rm 152}$,
R.~Abreu$^{\rm 116}$
% ...
\bigskip
\\
% ...
$^{11}$ Institute of Physics, Azerbaijan Academy of Sciences, Baku, Azerbaijan\\
% ...
$^{85}$ CPPM, Aix-Marseille Universit{\'e} and CNRS/IN2P3, Marseille, France\\
% ...
$^{90}$ Department of Physics and Astronomy, Michigan State University, East Lansing MI, United States of America\\
% ...
$^{107}$ Nikhef National Institute for Subatomic Physics and University of Amsterdam, Amsterdam, Netherlands\\
% ...
$^{113}$ Homer L. Dodge Department of Physics and Astronomy, University of Oklahoma, Norman OK, United States of America\\
% ...
$^{116}$ Center for High Energy Physics, University of Oregon, Eugene OR, United States of America\\
% ...
$^{151}$ Institute of Physics, Academia Sinica, Taipei, Taiwan\\
$^{152}$ Department of Physics, Technion: Israel Institute of Technology, Haifa, Israel\\
$^{153}$ Raymond and Beverly Sackler School of Physics and Astronomy, Tel Aviv University, Tel Aviv, Israel\\
% ...
$^{158}$ Department of Physics, University of Toronto, Toronto ON, Canada\\
% ...
\end{flushleft}

\clearpage

\end{document}